\documentclass[conference]{IEEEtran}
\IEEEoverridecommandlockouts
\usepackage{cite}
\usepackage{amsmath,amssymb,amsfonts}
\usepackage{mathtools}
\usepackage{algorithmic}
\usepackage{graphicx}
\usepackage{textcomp}
\usepackage{xcolor}
\usepackage{soul}
\usepackage{booktabs}
\usepackage{amsmath}
\usepackage{caption}
\usepackage{subcaption}
\usepackage{multirow}
\RequirePackage[bookmarksnumbered,unicode]{hyperref}

\usepackage[ruled,linesnumbered]{algorithm2e}
\SetKwInput{KwData}{Input}
\SetKwInput{KwResult}{Output}
\RestyleAlgo{ruled}

\def\BibTeX{{\rm B\kern-.05em{\sc i\kern-.025em b}\kern-.08em
    T\kern-.1667em\lower.7ex\hbox{E}\kern-.125emX}}

\begin{document}

\title{AIM: A practical approach to automated index management for SQL databases}

\author{\IEEEauthorblockN{Ritwik Yadav}
\IEEEauthorblockA{\textit{Database Engineering} \\
\textit{Meta Platforms, Inc.}\\
Menlo Park, USA \\
ritwikyadav@fb.com}
\and
\IEEEauthorblockN{Satyanarayana R. Valluri}
\IEEEauthorblockA{\textit{Database Engineering} \\
\textit{Meta Platforms, Inc.}\\
Menlo Park, USA \\
satyav@fb.com}
\and
\IEEEauthorblockN{Mohamed Za\"it}
\IEEEauthorblockA{\textit{Database Engineering} \\
\textit{Meta Platforms, Inc.}\\
Menlo Park, USA \\
mzait@fb.com}
}

\maketitle

\begin{abstract}
This paper describes AIM (Automatic Index Manager), a configurable index management system, which identifies impactful secondary indexes for SQL databases to efficiently use available resources such as CPU, I/O and storage. It has been validated on thousands of databases which support production systems. With AIM, the physical design of the database adapts itself to the changes in the workload.

We lay out the end to end design of AIM while calling out the guarantees and tradeoffs associated with our design choices. Some of the salient features of AIM include fast convergence even while recommending wide composite indexes, reduced reliance on the query optimizer and a “no regression” guarantee for production workloads. Each index recommendation from AIM is accompanied with a metrics driven explanation, making it easier to verify machine driven changes.

AIM is one of the few industrial strength index recommendation engines that is deployed on production databases at a large scale. The experimental results show that AIM is quick in identifying the most effective indexes and the resulting physical design is close to optimal.
\end{abstract}

\begin{IEEEkeywords}
automatic indexing, self managing databases, workload adaptivity
\end{IEEEkeywords}

\section{Introduction}
\subsection{Motivation}

A significant challenge in managing databases is the maintenance of efficient physical structures to facilitate data access for evolving workloads. Historically, this was done manually by database administrators with significant domain expertise. However, the emergence of cloud managed databases and pay-as-you-go models has resulted in automation of database tuning processes. The resulting decline in human involvement has made physical design tuning significantly more efficient at scale.

A carefully selected set of secondary indexes can have a defining impact on the applications powered by relational databases. The tradeoff involves using more storage in exchange for reduced CPU and I/O utilization. Reduced I/O during query execution vastly improves its latency. Decades of research \cite{agrawal2005database,bruno2007online,chaudhuri1997efficient,chaudhuri1998autoadmin,chaudhuri2007self,dageville2004automatic,valentin2000db2,zilio2004db2} has explored a variety of ways to navigate and choose from the space of available indexes. However, very few industrial strength solutions exist today that have been validated on a variety of production workloads and are capable of identifying wide secondary indexes in a \emph{reasonable} amount of time.

\subsection{Challenge}

Selection of the optimal set of indexes for a given workload is an NP-hard problem \cite{chaudhuri2004index}. In fact, optimal selection of indexes for facilitating a single join query is a complex problem in itself since determining the optimal join order is an NP-hard problem as well \cite{leis2015good}. 

Most state-of-the-art solutions \cite{schlosser2019efficient, dta} rely on utilizing the query optimizer to evaluate the cost of executing queries with hypothetical configurations of indexes. The number of these configurations explodes for real-life production workloads placing significant restraints on the solution space \cite{kossmann2020magic}. These restraints limit the ability of modern algorithms to identify wide composite indexes \emph{in time} and prevent severe performance degradation for input workloads. Prior work \cite{papadomanolakis2007efficient} also suggests that there is significant utility in optimizing the evaluation of these hypothetical configurations.

\subsection{Approach \& Contribution}

In this paper, we approach automatic index management from a practical viewpoint. Our design falls back on first priniciples of the utility of indexes. The benefit of using an index always results from reduced I/O operations. It can be attributed to efficient selection of records, reading only relevant attributes of selected records and / or by eliminating the need to sort selected records. The distinguishing contributions of our work are as follows:

\begin{itemize}
    \item Utilization of the \emph{structural properties} of individual queries to significantly restrict the search space of candidate indexes. The query structures are used to generate targeted index candidates rather than attempt to explore the entire space.
    \item Efficient exploration of \emph{wide multi-column indexes} becomes more feasible. Most modern techniques have to restrict the index width \cite{kossmann2020magic} in order to prevent the number of index configurations from exploding.
    \item To the best of our knowledge, AIM is one of the few widely deployed production algorithms which treats \emph{complex join queries} systematically.
    \item \emph{Reduced reliance} on the database query optimizer. Unlike its predecessors, AIM does not query the optimizer for data distribution statistics every time it has to append a new column to a multi-column candidate index. This is reflected in AIM's runtime which is orders of magnitude smaller than other modern algorithms.
    \item Our experimental study shows that AIM can both fill the gap left by manual tuning and build secondary indexes from scratch with comparable (or better) performance. It can also detect and drop (parts of) unused indexes.
\end{itemize}

\subsection{Structure of the paper}

Section \ref{section:problem} defines the index tuning problem which is followed by a detailed description of the general algorithm and system architecture used for solving the problem in ~\autoref{section:algo} and ~\autoref{section:candigen}. The algorithm described is general enough to be implemented by any SQL database and its salient features are called out in ~\autoref{section:salfet}. Conclusions drawn from experiments conducted on production workloads and synthetic benchmarks are presented in ~\autoref{section:expt}, which demonstrate the impact of the design choices made when implementing AIM. It also compares AIM against other modern algorithms and highlights the significantly better runtime of AIM. External components that play a crucial role in AIM's architecture are described in ~\autoref{section:support_comp}. Section \ref{section:lessons} calls out the lessons learned in the process of implementing AIM while ~\autoref{section:relatedw} draws a comparison with existing work and lists ideas for future work.

\section{Problem Definition}
\label{section:problem}

The index tuning problem focuses on selecting an optimal configuration $\mathcal{X}$ for a relational database $D$ such that the execution cost of the workload running on it ($\mathcal{W}$) can be minimized under certain constraints. For the scope of this problem, the configuration $\mathcal{X}$ is comprised of a set of secondary indexes, each of which is materialized on one of the tables present in $D$. The workload $\mathcal{W}$ is comprised of queries running on $D$. Each query, $q$, has an associated weight $w_q$. $w_q$ can be based on query $q$'s execution frequency, its overall CPU consumption or a manually specified importance measure provided by query authors.

\subsection{Static Index Tuning Problem}

The static index tuning problem (also referred to as bootstrapping) involves finding the optimal subset of possible secondary indexes that minimize the computational cost of serving a constant workload. It is formally defined as follows. 

Given a constant workload $\mathcal{W}$ (consisting of queries $q \in \mathcal{W}$), a set of all possible indexes $\mathcal{U}$ and a storage budget $B$, find $\mathcal{X^*}$ such that 

\begin{equation}
\label{eq:staticopt}
\mathcal{X^*} = \underset{\mathcal{X} \in 2^\mathcal{U}}{\operatorname*{argmin}} (\underset{q \in \mathcal{W}}{\sum} w_{q}cost(q, \mathcal{X}))
\end{equation}

where $w_q$ is the weight associated with query $q$ and $cost(q, \mathcal{X})$ is the execution cost of $q$ with index configuration $\mathcal{X}$ such that $\mathcal{X}$ fits in $B$. 

\subsection{Continuous Index Tuning Problem}
The continuous index tuning problem is a slightly different variant of the bootstrapping problem. In fact, this variant is closer to the real world since most developers deploy their databases with an initial set of indexes. This initial set might not always be the most efficient choice but it serves as a starting point to improve upon. The problem statement is as follows.

Given a workload $\mathcal{W}$ (consisting of queries $q \in \mathcal{W}$), a set of pre-existing indexes $\mathcal{C}$ and a storage budget $B$, find $\mathcal{C'}$ such that

\begin{equation}
\label{eq:overall_cost}
\underset{q \in \mathcal{W}}{\sum} w_{q}cost(q, \mathcal{C'}) \leq (1+\lambda_{1})\underset{q \in \mathcal{W}}{\sum} w_{q}cost(q, \mathcal{X^*})
\end{equation}

\begin{equation}
\label{eq:improve}
{\exists} q \in \mathcal{W} \bullet w_{q}cost(q, \mathcal{C'}) \leq (1-\lambda_{2}) (w_{q}cost(q, \mathcal{C}))
\end{equation}

\begin{equation}
\label{eq:regress}
{\forall} q \in \mathcal{W} \bullet w_{q}cost(q, \mathcal{C'}) \leq (1+\lambda_{3}) (w_{q}cost(q, \mathcal{C}))
\end{equation}

where $0 \leq \lambda_1, \lambda_2, \lambda_3 < 1$ are parameters that can be set to achieve the desired goals. Equation \ref{eq:overall_cost} keeps the overall cost in check with respect to the configuration $\mathcal{X*}$ detected while bootstrapping, equation \ref{eq:improve} mandates the improvement in performance of at least one query and equation \ref{eq:regress} prevents significant regressions in the performance of individual queries. 

The reason why databases need to undergo continuous tuning is because $w_q$ and $cost(q, \mathcal{X})$, for a given query $q$ and configuration $\mathcal{X}$, are functions of time as the workload and underlying data distributions change.

\section{Algorithm}
\label{section:algo}

\subsection{Terminology}
A detailed description of the algorithms requires defining some frequently used concepts for ease of understanding.

\subsubsection{Normalized query}
A normalized (or parameterized) form of a query is obtained by replacing parameters with placeholders. This mechanism is used to group together queries with similar structure. Note the $?$ used to represent a parameter in the following example. 

\begin{small}
\begin{verbatim}
 SELECT id, name FROM students WHERE score > ?
\end{verbatim}
\end{small}

\subsubsection{Discarded data ratio (ddr)}
For every execution of a query, we compute the amount of data that was read during its execution and the amount of data that was returned by it. The disparity between these two metrics per execution of a normalized query is computed as the ratio of data sent to data read averaged across executions of a normalized query.

\subsubsection{Partial order of index columns}
Potential index candidates are denoted as strict partial orders \cite{dushnik1941partially} of columns on a single table such as $(X_t, \prec)$ where $X_t$ is the set of columns of table $t$ and the relation $\prec$ denotes that a column precedes another in an index. Consider the following partial order where its ordered partitions are listed.

\begin{small}
\begin{verbatim}
   <{col1, col2}, {col3}, {col5, col6, col7}>
\end{verbatim}
\end{small}

It represents all non-unique secondary indexes where columns $col1$ and $col2$ occupy the first two places (order immaterial), followed by column $col3$ in third place. $col3$ is, in turn, followed by any permutation of $col5, col6,$ and $col7$ in the last three places.

\subsubsection{Dataless indexes}
\label{subsubsec:dataless}
Dataless indexes are secondary indexes that are created without actual data. They only have data distribution statistics and are used for estimating query execution costs by the optimizer\footnote{The estimated execution costs might be a bit off since index dives cannot be performed}. Dataless indexes are similar in implementation to “what-if” indexes described by Chaudhuri and Narasayya \cite{chaudhuri1998autoadmin}.

\subsubsection{Covering index}
\label{section:covering}
A covering index is a secondary index that constitutes all the columns of a table that are referenced in a specific query. Therefore, the clustered primary key (base table) need not be accessed in order to compute the result of the query. Covering indexes usually reduce the number of random disk seek operations.

\subsection{Overview}
\begin{figure}[htbp]
  \centering
  \includegraphics[width=\linewidth]{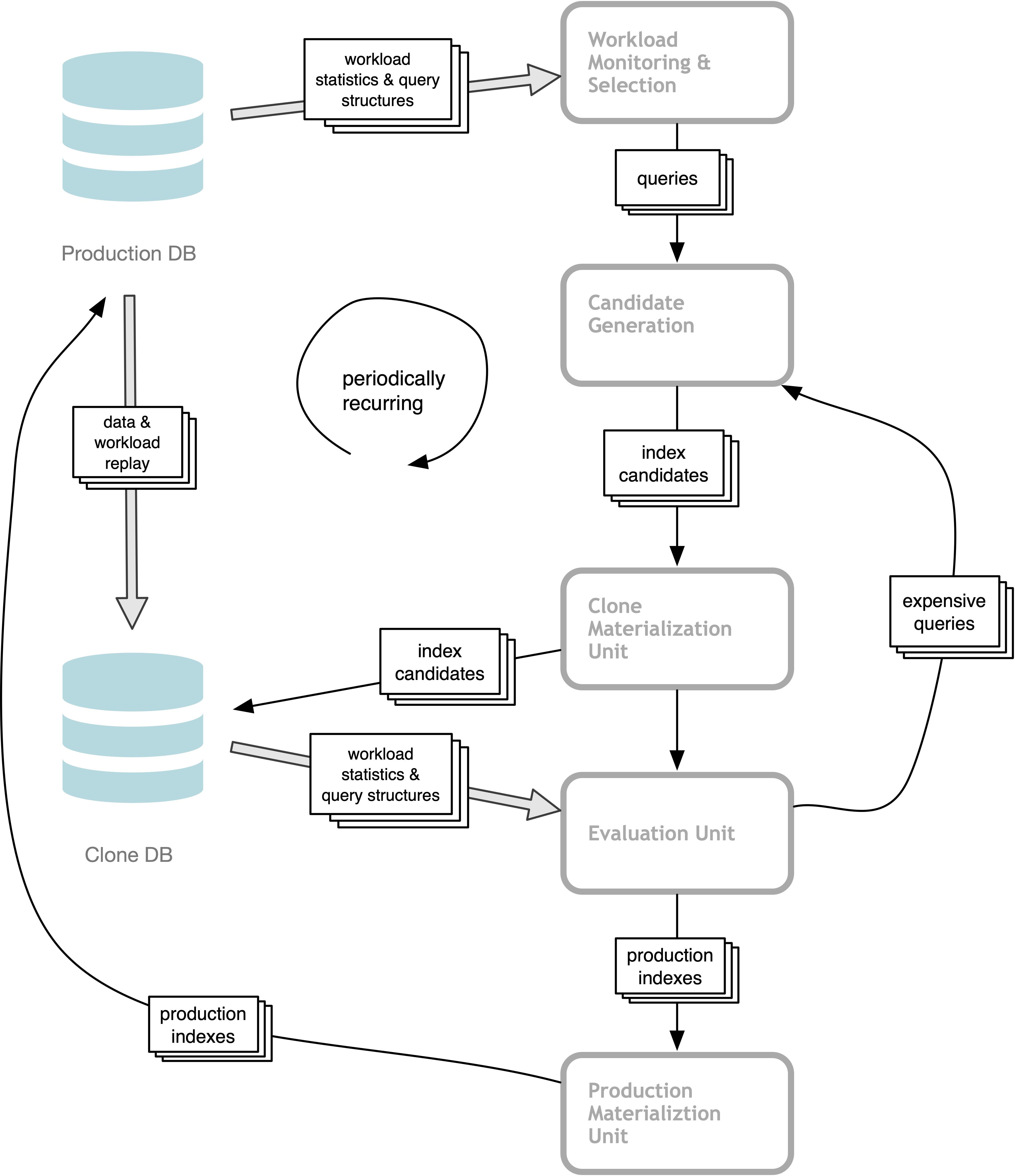}
  \caption{AIM flowchart}
  \label{fig:algo}
\end{figure}

The high-level flow of the AIM procedure is depicted in ~\autoref{fig:algo}. The algorithm can be thought of as having two distinct phases; both of which follow the same general outline presented in ~\autoref{alg:new_main}.
Initially, if multiple queries are executing inefficiently, it is not easy to predict their relative frequency in steady state after adding the necessary supporting indexes. Therefore, AIM tries to add indexes of smaller width to support each inefficient query in the observed workload. In a subsequent phase, it identifies queries that might benefit from covering indexes. Usually, these queries execute extremely frequently to offset the extra storage overhead of creating wider indexes.

\begin{algorithm}[htbp]
\small
\caption{Automatic Index Manager}\label{alg:new_main}
\KwData{$database$: name of the database, $j$: join parameter}
\KwResult{$production\_indexes$: Set of index to be created in production on $database$}
    $\mathcal{W} \gets WorkloadSelection(database)$\;
    $candidates \gets GenerateCandidates(\mathcal{W}, j)$\;
    Materialize $candidates$ on the clone $database$, ordering them in descending order of perceived benefits from optimizing queries in $\mathcal{W}$ until storage budget is exhausted\;
    $production\_indexes \gets RankSelectedIndexes(candidates)$
\end{algorithm}

Line 1 of ~\autoref{alg:new_main} denotes the representative workload selection step which identifies the set of queries that need tuning by monitoring their execution statistics. For each query that is identified by the previous step, a set of candidate indexes are generated (line 2). The join parameter $j$ determines which tables (out of the many) present in a complex join query are exhaustively evaluated for all possible relative join orders (explained in ~\autoref{sub:joins}). 

The set of candidate indexes is materialized on a clone of the database to determine if the query optimizer is able to use the indexes efficiently during execution (line 3). This is necessary to guarantee that none of the queries regress in production (more details in ~\autoref{section:clone}). The set of production indexes is identified by ranking the usefulness of the candidate indexes (line 4) which are then materialized on the production database. 

The algorithm works for both static and continuous index tuning. It can be run once to bootstrap the initial set of indexes when the database is created and periodically thereafter for continuous tuning to adapt to changes in the workload. The following sub-sections provide detailed description of each of the steps outlined in the high level algorithm.

\subsection{Representative workload selection}
\label{subsection:workloadsel}
The goal of this step is to identify the set of queries that require index tuning. The workload monitor analyzes execution statistics to identify queries that are not being executed efficiently. Query execution statistics include important information such as CPU cost, rows read, rows sent and number of executions corresponding to each normalized query. The utility of this data is in selecting / ordering queries with maximum expected benefits from adding potentially useful indexes.

Based on the query execution statistics, a representative sample of the workload is selected by identifying the most expensive queries that need tuning. The selection of the representative workload for a given database is fully automated and done periodically at the end of a configurable interval of time. The selection process takes three main attributes into account for each query; execution frequency, average CPU consumption per execution and the discarded data ratio.

The threshold on the execution frequency is only used to weed out spurious executions of queries from ad hoc sources. The combination of average CPU consumption and discarded data ratio ($ddr_{avg}$) is used to assign an optimistic expected benefit ($B$) from optimizing a normalized query $q$ executed with configuration $\mathcal{X}$. 

\begin{equation}
\label{eq:wseleq}
B(q, \mathcal{X}, \Delta t) \coloneqq (1 - ddr_{avg}(q, \mathcal{X}, \Delta t)) \cdot cpu_{avg}(q, \mathcal{X}, \Delta t)
\end{equation}

$cpu_{avg}(q, \mathcal{X}, \Delta t)$ represents the average CPU utilization by normalized query $q$ over the time span $\Delta t$. It is measured in CPU seconds and includes cycles spent on $CPU\_IOWAIT$ events. ~\autoref{eq:wseleq} assumes that all the I/O done by the query which isn't returned in the result set could have been avoided by proper index structures and thereby represents the maximum potential benefit from optimizing the query\footnote{This assumption is not always true for certain cases such as queries with a GROUP BY operation}. A threshold on $B$ (e.g. 1/20 of a CPU core) is used to select the queries in the workload targeted for optimization.

\subsection{Generate Candidate Indexes}
\label{sub:gci}

The candidate generation process takes into account the structure of the query and is \emph{efficiently} able to generate effective composite indexes. Statistics around data distribution are indirectly taken into account in our implementation via dataless indexes. The key lies in defining transformation functions from column usage metadata to potential index candidates represented by a partial order of columns. The basic transformation functions are more or less the same across most popular SQL database servers and this approach reduces the number of candidates considerably.

\begin{algorithm}[htbp]
\small
\caption{GenerateCandidates}
\label{alg:candgen}
\KwData{$\mathcal W$: workload, $j$: join parameter}
\KwResult{$\mathcal {IP}$: Set of candidate indexes}
$\mathcal{PO} \gets \phi$\;
\ForEach{$\mathcal{Q} \in \mathcal W$}{
    $mode \gets TryCoveringIndex(\mathcal{Q})$\;
    $\mathcal{PO} \gets \mathcal{PO} \cup GenerateCandidatesForSelection(\mathcal Q, j, mode) \cup GenerateCandidatesForGroupBy(\mathcal Q, j, mode) \cup GenerateCandidatesForOrderBy(\mathcal Q, j, mode)$\;
}
$\mathcal{PO}_{final} \gets MergePartialOrders(\mathcal{PO})$\;
return $GenerateCandidateIndexPerPO(\mathcal{PO}_{final})$\;
\end{algorithm}

Algorithm \ref{alg:candgen} shows the high level algorithm for the generation of candidate indexes. The algorithm takes as input the workload and the join parameter $j$. For each query $\mathcal{Q}$ in the workload, the algorithm checks if a covering index would be beneficial for it (line 3). The $TryCoveringIndex$ procedure makes this determination based on the indexes utilized for the execution of $\mathcal{Q}$ in the current configuration. A covering index is tried if it is not possible to improve selectivity any further and the addition of a covering index is expected to significantly bring down the number of additional seeks from the primary key. Taking the example of Q4 in ~\autoref{sub:gby}, a covering index is not tried until an index with the prefix $<$col2, col3$>$ is already being utilized. Furthermore, if such an index is already being used, the number of disk seeks should be high enough to offset the cost of paying the extra storage cost of including one more column $col1$ to the index. This threshold is high for fast storage media such as SSDs. Line 4 of the algorithm generates a set of partial orders of index columns by taking into consideration alternative query execution strategies. The execution strategy is based on the primary operation (out of selection, grouping or ordering) that is optimal for the query. The primary operation is the one that precedes all other operations. For instance, one execution strategy of query $\mathcal{Q}$ might involve sorting rows in group order before selection of rows, whereas another strategy might prefer selection before sorting in group order.

The set of partial orders generated in the previous steps are merged to get the final set of partial orders (line 6). The procedure is described in detail in ~\autoref{subsec:merge}. Corresponding to each partial order, one index candidate is generated by arbitrarily choosing a total ordering of index columns which satisfies the partial order. This is done by the $GenerateCandidateIndexPerPO$ procedure (line 7).

\subsection{Merge Partial Orders}
\label{subsec:merge}

We first define a function $MergeCandidatesPairwise$ which takes two strict partial orders $(P, \prec_P)$ and $(Q, \prec_Q)$ to produce another strict partial order $(R, \prec_R)$ when the prerequisite condition $C_{merge}$ is met.
\[
C_{merge} \coloneqq P \subseteq Q \bigwedge (\nexists \text{ } a, b \in P : a \prec_P b \wedge b \prec_Q a)
\]
\[
(R, \prec_R) \coloneqq \left\{
\begin{array}{lc}
    (P, \prec_P) \bigoplus (Q, \prec_Q), & \text{if } C_{merge}\\
    (\phi, \prec_R), & \text{otherwise}
\end{array}
\right.
\]
$P$, $Q$ and $R$ are subsets of columns of the same table. $a \prec_P b$, $a \prec_Q b$ and $a \prec_R b$ denote relations; all of which semantically translate to the fact that column $a$ precedes column $b$ in the index represented by these partial orders. $\bigoplus$ denotes the ordinal sum \cite{appelgate1969ordinal} operator.

The function $MergePartialOrders$ (from ~\autoref{alg:candgen}) takes a set of partial orders of index columns ($\mathcal{PO}$) and recursively merges the constituent partial orders in pairs (using $MergeCandidatesPairwise$) until no new ones are produced. The following equation give a more formal definition.
\begin{multline}
\label{eq:mpo1}
\mathcal{PO}_{n+1} \coloneqq \\ \{MergeCandidatesPairwise(X, Y) \mid X, Y \in {PO}_{n} \}
\end{multline}

$MergePartialOrders(\mathcal{PO})$ starts with $\mathcal{PO}_0 \coloneqq \mathcal{PO}$ and returns $\mathcal{PO}_{m}$ using ~\autoref{eq:mpo1} such that $\mathcal{PO}_{m} = \mathcal{PO}_{m + 1}$. Intuitively, merging of the partial orders of index columns helps in picking the right order of index keys that will maximize the benefit of the index created. For example, consider the following two partial orders: $<$\{col1, col2, col3\}$>$ \& $<$\{col2, col3\}$>$.

The first partial order denotes that some query might benefit from the creation of an index with any permutation of $col1, col2$ and $col3$. The second partial order denotes that some query might benefit from the creation of an index with any permutation of $col2$ and $col3$. Merging the partial orders leads to creation of a new partial order: $<$\{col2, col3\}, \{col1\}$>$.

The merged partial order puts a constraint on the index columns' ordering to be limited to $<$col2, col3, col1$>$ and $<$col3, col2, col1$>$. Either candidate satisfies the merged partial orders and can individually be beneficial to queries for which the base partial orders were merged.

\subsection{Ranking and selection of candidates}
\label{subsec:rank}
The utility of an index is computed by summing up the benefits observed by individual queries that use it and discounting the write amplification resulting from index maintenance. The cost of a SQL query $q$ can be decomposed into two main components expresssed as $cost(q, \mathcal{X}) = cost_r(q, \mathcal{X}) + \sum_{i \in \mathcal{X}}cost_u(q, i)$ where $cost_r(q, \mathcal{X})$ represents the cost of locating the records by using indexes in configuration $\mathcal{X}$ and $cost_u(q, i)$ represents the overhead of updating index $i$ in $\mathcal{X}$. $cost_u(q, i)$ is non-zero only for DML statements. The logic for estimating (components of) costs is dependent on the storage engine \cite{costem} and can utilize the statistics offered by dataless indexes. 

The two main ways in which indexes help queries are by reducing the amount of data that needs to be read and preventing repeated sorting operations. In order to quantify this benefit, every query $q$ is treated in isolation at first. This gain ($U_+$) can be estimated by ~\autoref{eq:iben} where $\mathcal{I}$ is the set of candidates generated to benefit query $q$. $U_+$ is distributed amongst the indexes (in $\mathcal{I}$) which are used during $q$'s execution. The share $s_{i,q}$ of $U_+$ attributed to an index $i$ is directly proportional to the reduction in I/O due to $i$ (estimated by the optimizer).

\begin{equation}
\label{eq:iben}
U_+(q, \mathcal{I}) \coloneqq \frac{cost(q, \phi) - cost(q, \mathcal{I})}{cost(q, \phi)} \cdot cpu_{avg}(q, \phi, \Delta t)
\end{equation}

Similarly, the overhead of maintenance arises from index updates and storage space required. The overhead of index maintenance can be computed for an individual index $i$ using ~\autoref{eq:ilos}. Higher I/O operations lead to higher CPU utilization because the $cpu_{avg}$ metric includes $CPU\_IOWAIT$ events. 

\begin{equation}
\label{eq:ilos}
u_-(i) \coloneqq \underset{q \in \mathcal{W}}{\sum} \frac{cost_u(q, i)}{cost(q, \phi)} \cdot cpu_{avg}(q, \phi, \Delta t)
\end{equation}

The overall utility of an index is represented by $u(i) = s_{i,q} \cdot U_+(q, \mathcal{I}) + u_-(i)$. The accounting for index interactions is limited to the merging of representative partial orders of the index candidates. When index candidates are merged, the benefits corresponding to individual queries gets added up and write amplification overhead is the same as that of the wider candidate being merged. Index selection can then be modeled as a knapsack problem \cite{martello1987algorithms} where index candidates are evaluated in the order of their overall utility per unit storage overhead while not violating the budget allocated for indexes. We implemented several other heuristics for a more complete accounting of index interactions but comparison of their relative effectiveness is deferred to future work.

\section{Candidate generation}
\label{section:candigen}

Candidate generation takes into account the structure of the query and transforms structural metadata to partial orders representing index candidates. The structural metadata collected by the workload monitor includes information about operations corresponding to each column, edges in the table join graph and factors in the selection predicate. Examples of this metadata are listed in \autoref{tab:columeta}. This information is particularly useful in generating candidate indexes for common SQL operations as is described in the following subsections.

\begin{table*}[ht]
  \caption{Query structure}
  \label{tab:columeta}
  \centering
  \begin{tabular}{c|c|c}
    \multirow{2}{*}{Column Usage Metadata} & Operation & FILTER, TABLE JOIN, GROUP BY, ORDER BY, PROJECTION etc. \\
    & Operator & GREATER THAN, EQUAL, SORT ASCENDING etc.\\
    \midrule
    Structural metadata &\multicolumn{2}{c}{Edges in the table join graph, grouping of predicates in AND-OR chains etc.}\\
  \end{tabular}
\end{table*}

\subsection{Projection}
The projection operation is used to select a subset of the attributes (columns) from a relation (table). Let us assume the existence of a table $t_1$ with five columns, namely, $col1, col2, col3, col4$ and $col5$. A very simple example demonstrating projection would be the following.

\begin{small}
\begin{verbatim}
  Q1: SELECT col2, col3 FROM t1 WHERE col5 < 2
\end{verbatim}
\end{small}

In this query, instead of requesting all columns from $t_1$ for rows that satisfy the WHERE clause, only $col2$ and $col3$ are requested in the output. The following partial order represents the set of indexes that would optimize Q1 in the example above: $<$\{col5\}, \{col2, col3\}$>$.

The consideration of these secondary indexes which prevent lookups to the primary key can be useful when using slow storage media. It should also be noted that addition of \emph{both} $col2$ and $col3$ at the end is necessary to avoid primary key lookups. Most greedy algorithms which add one column at a time may not be able to discover covering indexes for common queries.

\subsection{Selection (Filter operation)}
\label{section:selection}

The WHERE clause may contain predicates of the following form connected by logical operators (e.g. AND, OR)

\begin{small}
\begin{verbatim}
            column_name op expression
\end{verbatim}
\end{small}

These predicates which aid in the selection of rows from a single table instance are termed as filter operations. For each distinct table instance, we construct partial orders of columns that participate in filter operations. For e.g., a predicate like 

\begin{small}
\begin{verbatim}
    E1: WHERE col1 = 5 AND col2 = ‘ABC’ 
            AND col3 IN (5, 9, 11)
\end{verbatim}
\end{small}

would result in the following partial order of columns: $<$\{col1, col2, col3\}$>$.

\subsubsection{Complex AND-OR predicates}
The selection predicate might employ complex AND-OR chains and generation of candidates may differ slightly depending on the factorization technique implemented by the database server. This technique is denoted by the $FactorizeIndexPredicates$ routine in ~\autoref{alg:genc4}. The algorithm employed by MySQL is simple but other database engines may use more complex techniques of factorization \cite{chaudhuri2003factorizing}. In our implementation, we simply use the disjunctive normal form (DNF) for complex predicates and it works well with MySQL. Each factor in the DNF results in a separate partial order. For e.g., the following predicate results in two partial orders; $<$\{col1, col2, col3\}$>$ and $<$\{col2, col4\}$>$.

\begin{small}
\begin{verbatim}
    E2: WHERE (
            col1 = 5 AND col2 = ‘ABC’ 
                AND col3 IN (5, 9, 11)
        ) OR ( col2 = ‘CDE’ AND col4 = 8 )
\end{verbatim}
\end{small}

\subsubsection{Index prefix predicates}
Depending on the database server, the operator plays an important role as well. Filtered upon columns associated with operators like equal ($=$), nullsafe equal ($<=>$), set membership etc. can be chained together to construct multi-column indexes wherein each participating column would typically enhance the selectivity (or the effectiveness) of the index. However, if columns associated with operators such as greater than ($>$), less than or equal to ($<=$), $BETWEEN$ etc. were concatenated together to form a multi-column secondary index, each subsequent column after the first one may not have a strict \emph{additive} benefit on selectivity \cite{multirange}. This is best explained with an example.

\begin{small}
\begin{verbatim}
    E3: WHERE col1 = 5 AND col2 = ‘ABC’ 
            AND col3 > 5 AND col4 < 2.0
\end{verbatim}
\end{small}

If a multi-column secondary index on $<$col1, col2, col3, col4$>$ is added, all rows with $col1 = 5$, $col2 = ABC$ and $col3 > 5$ might have to be read and the evaluation of $col4 < 2.0$ would be done in a subsequent step. This subsequent step maybe performed by reading the qualifying rows from primary key or by using the index condition pushdown optimization \cite{icp}. Therefore, we can construct the following partial order for the expression E3: $<$\{col1, col2\}, \{col3, col4\}$>$. 

This partial order represents a set of 4 multi-column secondary indexes where $col1$ and $col2$ precede $col3$ and $col4$. The order of $col1$ with respect to $col2$ is immaterial. Similarly, the order of $col3$ with respect to $col4$ would have been immaterial as well but it depends on the overall selectivity of the individual predicates $col3 > 5$ and $col4 < 2.0$. Putting the column corresponding to the more selective atomic predicate first would be more efficient and can be done by using dataless indexes. Therefore, the subpredicates $col1=5$ and $col2=$\textquoteleft$ABC$\textquoteright$\ $decide the prefix of the candidate index. This process is described by the procedure $GenerateCandidateIndexPredicates$ in ~\autoref{alg:genc4}.

More formally, an index prefix predicate is an atomic subpredicate of the filter clause which can be resolved by an index scan of rows that share a constant non-empty prefix. This constant prefix can be determined from the atomic subpredicate itself. For e.g. the atomic predicate $col1 = 5$ can be resolved using an index on $col1$ by scanning all the rows that have the prefix 5 and $col2 = $\textquoteleft$ABC$\textquoteright$\ $can be resolved using an index on $col2$ by scanning all the rows with prefix \textquoteleft$ABC$\textquoteright. Contrary to this, the atomic predicate $col3 > 5$ can also be resolved using an index on $col3$ by performing a range scan. However, all the matching rows need \emph{not} have a common prefix. Therefore, $col3 > 5$ is not an index prefix predicate.

\subsection{Joins}
\label{sub:joins}
Table join operations compute a subset of the cartesian product of two or more tables (relations). Therefore, it can be thought of as selection operation (described in ~\autoref{section:selection}) on a cross-product of two or more tables.

For most database execution engines, only two tables are joined together at a time. The order in which tables are joined together plays an important role in determining the performance of a join query\footnote{The order of some join types is predetermined (e.g. straight join \cite{stjoin})}. Determining the most efficient join order for a query is in itself a NP hard problem \cite{toshihide1984optimal}. There is a circular dependency between the problem of selecting the most efficient indexes for facilitating a join query and determining the best join order for it. Most relational database engines employ heuristics to predict an efficient join order rather than computing the most efficient one. The goal is to prevent query optimization from becoming more expensive than actual execution. Therefore, only a small number of join orders are even considered by the optimizer \cite{haritsa2005analyzing}.

\begin{figure}[htbp]
  \centering
  \includegraphics[width=0.5\linewidth]{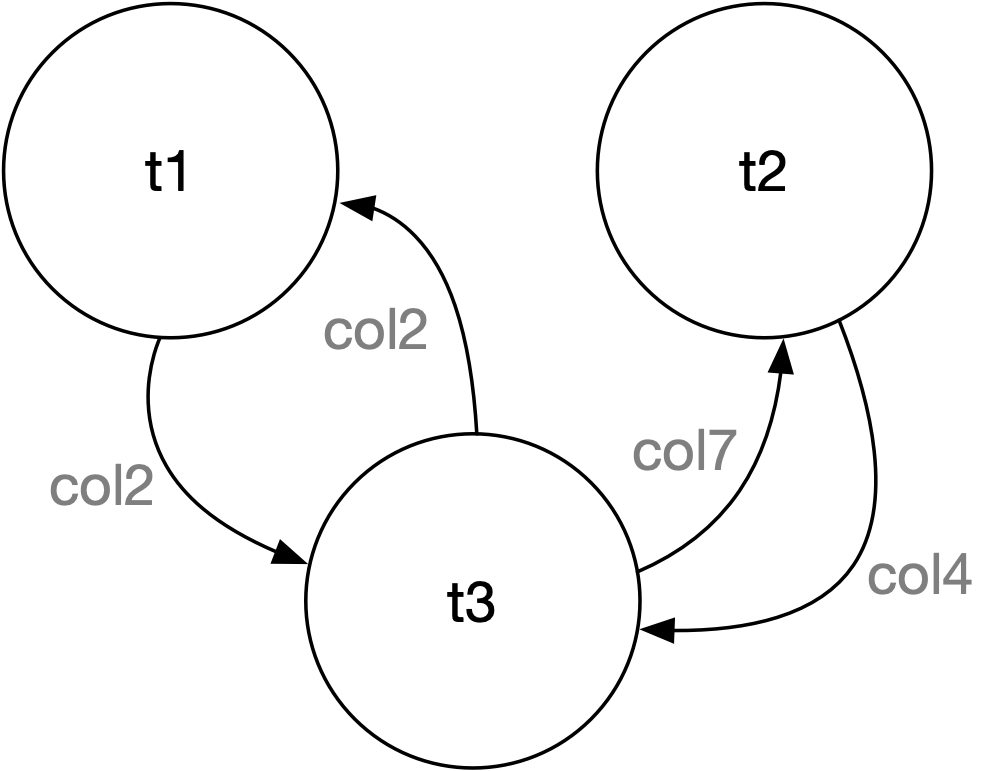}
  \caption{Join graph for Q2}
  \label{fig:jg1}
\end{figure}

Our approach can generate candidate indexes for join queries which works well for transactional workloads. Column usage metadata includes identifiers for table instances that participate in a join condition with any given column of another table instance. This helps us construct the join graph where nodes are table instances and edges between them represent a predicate in the join condition involving columns from the participating tables instances (nodes). Consider the following query Q2 with the join graph shown in ~\autoref{fig:jg1}.

\begin{small}
\begin{verbatim}
    Q2: SELECT t1.col1, t2.col2, t3.col3 
        FROM t1, t2, t3 
        WHERE t1.col2 = t3.col2 
            AND t2.col4 = t3.col7
\end{verbatim}
\end{small}

AIM accepts a positive integer $j$ to limit its search for join orders. All table instances with columns participating in the join predicates with at most $j$ other table instances are selected. The candidate indexes for the selected table instances include all possibilities of join orders with respect to that table. The upper bound is exponential in $j$, since every participating table could either precede or succeed the selected table instance. Therefore, the value of $j$ is usually a small positive integer.  Although, we do not try to exhaustively facilitate all join orders for table instances that share a join predicate with more than $j$ tables, we have not seen any incremental benefit of exploring $j > 3$ for any production workload so far. The impact of choosing $j$ is demonstrated in ~\autoref{fig:gph4}.

If the join order is predetermined, the join predicates can be merged with the selection predicates to generate partial orders representing candidate indexes benefiting both the join and selection operations. Intuitively, if we assume that the join order for the query Q2 above is [$t3$, $t1$, $t2$], it makes sense to try out an index on $t2.col4$ since $t3$ precedes it and we perform a lookup into table $t2$ for every value of $t3.col7$. However, if the join order were [$t1$, $t2$, $t3$], an index on $t2.col4$ would not reduce the number of records considered from $t2$. 

Our approach considers candidate indexes to facilitate several possible join orders and lets the query optimizer pick the best one. The pseudocode for generation of index candidates catering to filter and join predicates is provided in ~\autoref{alg:genc1}. The procedure $ReferencedColumns$ is an overloaded helper which returns the columns referenced in an entity. The entity could be a table in a query or a partial order.

It should again be noted that the greedy approach (utilized by most modern index selection algorithms) would not be able to discover optimal indexes for complex join queries because their exploration logic doesn't consider co-ordinated exploration of candidate indexes across multiple tables. 

\begin{algorithm}[htbp]
\small
\caption{JoinedTablesPowerset}\label{alg:powjoins}
\KwData{$\mathcal Q$: query, $t$: table, $j$: join parameter}
\KwResult{$\mathcal T_j$: Power set of tables which have join predicates with $t$}
$\mathcal{T} \gets$ Set of tables with join predicates with $t$ in $\mathcal{Q}$\;
\If{$|\mathcal{T}| > j$}{$\mathcal{T} \gets \phi$}
return $2^\mathcal{T}$
\end{algorithm}

\begin{algorithm}[htbp]
\small
\caption{GenerateCandidatesForSelection}\label{alg:genc1}
\KwData{$\mathcal Q$: query, $j$: join parameter, $mode$: covering / non-covering}
\KwResult{$\mathcal {IP}$: Set of partial orders of index columns}
$\mathcal {IP} \gets \phi$\;
$\mathcal T \gets$ tables queried in $\mathcal Q$ \;
\ForEach {$t \in \mathcal T $} {
    $\mathcal{C}_F \gets$ Set of columns in $t$ that feature in filter predicates of $\mathcal{Q}$\;
    \ForEach {$S \in JoinedTablesPowerset(\mathcal Q, t, j) $} {
        $\mathcal{C}_J \gets$ Set of columns in $t$ that feature in join predicates (of $\mathcal{Q}$) with any table in $S$\;
        $candidates \gets GenerateCandidateIndexPredicates(\mathcal Q, \mathcal{C}_F \cup \mathcal{C}_J)$\;
        \If {$mode = covering$} {
            $candidates \gets \{c.append(ReferencedColumns(\mathcal Q, t) \setminus ReferencedColumns(c)) \mid c \in candidates\}$\;
        }
        $\mathcal {IP} \gets \mathcal{IP} \cup candidates$\;
    }
}
return $\mathcal{IP}$
\end{algorithm}

\begin{algorithm}[htbp]
\small
\caption{GenerateCandidateIndexPredicates}
\label{alg:genc4}
\KwData{$\mathcal Q$: query, $\mathcal C$: columns used for selection}
\KwResult{$\mathcal {IP}$: Set of partial orders of index columns}
$\mathcal {IP} \gets \phi$\;
$\mathcal{G}_C \gets FactorizeIndexPredicates(\mathcal Q, \mathcal C)$\;
\ForEach {$\mathcal{G} \in \mathcal{G}_C $} {
    $\mathcal{C}_{IPP} \gets \{c \mid c \in \mathcal G$ and column $c$ features in an index prefix predicate (IPP)$ \}$\;
    $\mathcal{C}_{RSP} \gets \mathcal G \setminus \mathcal{C}_{IPP}$\;
    $last\_col \gets \operatorname*{argmin}_{c \in \mathcal{C}_{RSP}} dataless\_index\_cost(\mathcal Q, <\mathcal{C}_{IPP}, \{c\}>)$\;
    $\mathcal {IP} \gets \mathcal {IP} \cup \{<\mathcal{C}_{IPP}, \{last\_col\}>\}$\;
}
return $\mathcal{IP}$
\end{algorithm}

\subsection{Group By}
\label{sub:gby}
In case of a group by operation on a table instance, a secondary index can come in handy when reading rows in grouping order and computing any expression involving aggregation for each group. An example is as follows.

\begin{small}
\begin{verbatim}
Q3: SELECT col3, COUNT(*) FROM t1 
    GROUP BY col3
\end{verbatim}
\begin{verbatim}
Q4: SELECT col3, SUM(col1) 
    FROM t1 WHERE col2 = 5 GROUP BY col3
\end{verbatim}
\end{small}

For query Q3, a secondary index on $col3$ might come in handy when evaluating the aggregate values per group (defined by unique values of $col3$). For query Q4, however, the follwoing partial order would be more efficient since it forms a covering index while facilitating selection and grouping operations simultaneously: $<$\{col2\}, \{col3\}, \{col1\}$>$.

Notice that $col2$ precedes the grouping column ($col3$) since it features in an index prefix predicate defined above. Candidate generation for group by clause is described in ~\autoref{alg:genc2}.

\begin{algorithm}[htbp]
\small
\caption{GenerateCandidatesForGroupBy}
\label{alg:genc2}
\KwData{$\mathcal Q$: query, $j$: join parameter, $mode$: covering / non-covering}
\KwResult{$\mathcal {IP}$: Set of partial orders of index columns}
$\mathcal {IP} \gets \phi$\;
$\mathcal T \gets$ tables queried in $\mathcal Q$ \;
\ForEach {$t \in \mathcal T $} {
    $\mathcal{C}_G \gets$ Set of columns in $t$ that feature in the group-by clause within $\mathcal{Q}$\;
    
    \eIf{$mode = non-covering$}{
        $\mathcal{IP} \gets \mathcal{IP} \cup \{<\mathcal{C}_G>\}$\;
    }{
        $\mathcal{C}_F \gets$ Set of columns in $t$ that feature in filter predicates of $\mathcal{Q}$\;
        \ForEach {$S \in JoinedTablesPowerset(\mathcal Q, t, j) $} {
            $candidate \gets <>$\;
            $\mathcal{C}_J \gets$ Set of columns in $t$ that feature in join predicates (of $\mathcal{Q}$) with any table in $S$\;
            $\mathcal{G}_C \gets FactorizeIndexPredicates(\mathcal Q, \mathcal C_F \cup \mathcal C_J)$\;
            \ForEach {$\mathcal{G} \in \mathcal{G}_C $} {
                $\mathcal{C}_{IPP} \gets \{c \mid c \in \mathcal G$ and column $c$ features in an index prefix predicate (IPP)$ \}$\;
                $candidate \gets candidate.append(\mathcal{C}_{IPP})$\;
                $candidate \gets candidate.append(\mathcal{C}_G)$\;
                $remaining\_columns \gets ReferencedColumns(\mathcal Q, t) \setminus (\mathcal{C}_{IPP} \cup \mathcal{C}_G)$ \;
                $candidate \gets candidate.append(remaining\_columns)$\;
                $\mathcal{IP} \gets \mathcal{IP} \cup \{candidate\}$\;
            }
        }
    }
}
return $\mathcal{IP}$
\end{algorithm}

\subsection{Order By}
A SQL query might request only a few rows in the output which are sorted in a specific order. Consider the following query.

\begin{small}
\begin{verbatim}
    Q5: SELECT t1.col13, t1.col14, 
                    t2.col25, t1.col17 
        FROM t1 LEFT JOIN t2 
            ON t1.col11 = t2.col21        
        WHERE t1.col12 IN ('ABC', 'DEF') 
        ORDER BY t1.col13 LIMIT 2
\end{verbatim}
\end{small}

It is possible that the number of rows in $t_1$ matching the predicate t1.col12 IN ('ABC', 'DEF') is too high and a more efficient execution plan might read rows of $t_1$ in ascending order of $col13$ and check if the predicate t1.col12 IN ('ABC', 'DEF') is satisfied. In such a scenario, a secondary index on $col13$ would be more beneficial than one on $col12$. Candidate generation for order by clause is described by ~\autoref{alg:genc3}.

\begin{algorithm}[htbp]
\small
\caption{GenerateCandidatesForOrderBy}\label{alg:genc3}
\KwData{$\mathcal Q$: query, $j$: join parameter, $mode$: covering / non-covering}
\KwResult{$\mathcal {IP}$: Set of partial orders of index columns}
$\mathcal {IP} \gets \phi$\;
$\mathcal T \gets$ tables queried in $\mathcal Q$ \;
\ForEach {$t \in \mathcal T $} {
    $\mathcal{C}_O \gets$ Sequence of columns in $t$ that feature in the order-by clause within $\mathcal{Q}$\;
    
    \eIf{$mode = non-covering$}{
        $\mathcal{IP} \gets \mathcal{IP} \cup \{\mathcal{C}_O\}$\;
    }{
        $\mathcal{C}_F \gets$ Set of columns in $t$ that feature in filter predicates of $\mathcal{Q}$\;
        \ForEach {$S \in JoinedTablesPowerset(\mathcal Q, t, j) $} {
            $candidate \gets <>$\;
            $\mathcal{C}_J \gets$ Set of columns in $t$ that feature in join predicates (of $\mathcal{Q}$) with any table in $S$\;
            $\mathcal{G}_C \gets FactorizeIndexPredicates(\mathcal Q, \mathcal C_F \cup \mathcal C_J)$\;
            \ForEach {$\mathcal{G} \in \mathcal{G}_C $} {
                $\mathcal{C}_{IPP} \gets \{c \mid c \in \mathcal G$ and column $c$ features in an index prefix predicate (IPP)$ \}$\;
                $candidate \gets candidate.append(\mathcal{C}_{IPP})$\;
                $candidate \gets candidate.append(\mathcal{C}_{O})$\;
                $remaining\_columns \gets ReferencedColumns(\mathcal Q, t) \setminus (\mathcal{C}_{IPP} \cup \mathbf{elems}~\mathcal{C}_O)$\;
                $candidate \gets candidate.append(remaining\_columns)$\;
            }
            $\mathcal{IP} \gets \mathcal{IP} \cup \{candidate\}$\;
        }
    }
}
return $\mathcal{IP}$
\end{algorithm}

\section{Salient Features}
\label{section:salfet}
\subsection{Solution Granularity}
As described above, AIM compromises on the solution granularity to offer lower runtimes. Utilization of the query structure to severely limit our search space prevents the algorithm from exploring too many index configurations. It can support the following levels of solution granularity.
\begin{itemize}
    \item Only a subset of the queries can be chosen to be optimized. Anecdotally, only the top few most expensive queries account for most of the CPU utilization.
    \item At the query level, each table instance in a multi-table query has three options. It can either be unindexed, may have a non-covering index or a covering index.
    \item Relaxation / reduction of the number of sub-predicates in the index prefix predicates (IPP) can also lead to significant cost reductions especially when the additive selectivity falls below a certain threshold.
\end{itemize}

\subsection{Role of dataless indexes}
Dataless indexes (section ~\ref{subsubsec:dataless}) are only utilized when determining the join order (when a lot of tables are involved as described in ~\autoref{sub:joins}), the most selective atomic predicate that is not an index prefix predicate (~\autoref{alg:genc4}) and picking out the primary operation for each distinct query (~\autoref{sub:gci}). 

\section{Experimental Study}
\label{section:expt}
\subsection{Comparison with manual tuning}
\begin{figure*}[htbp]
    \begin{subfigure}[b]{0.325\textwidth}
        \includegraphics[width=\textwidth]{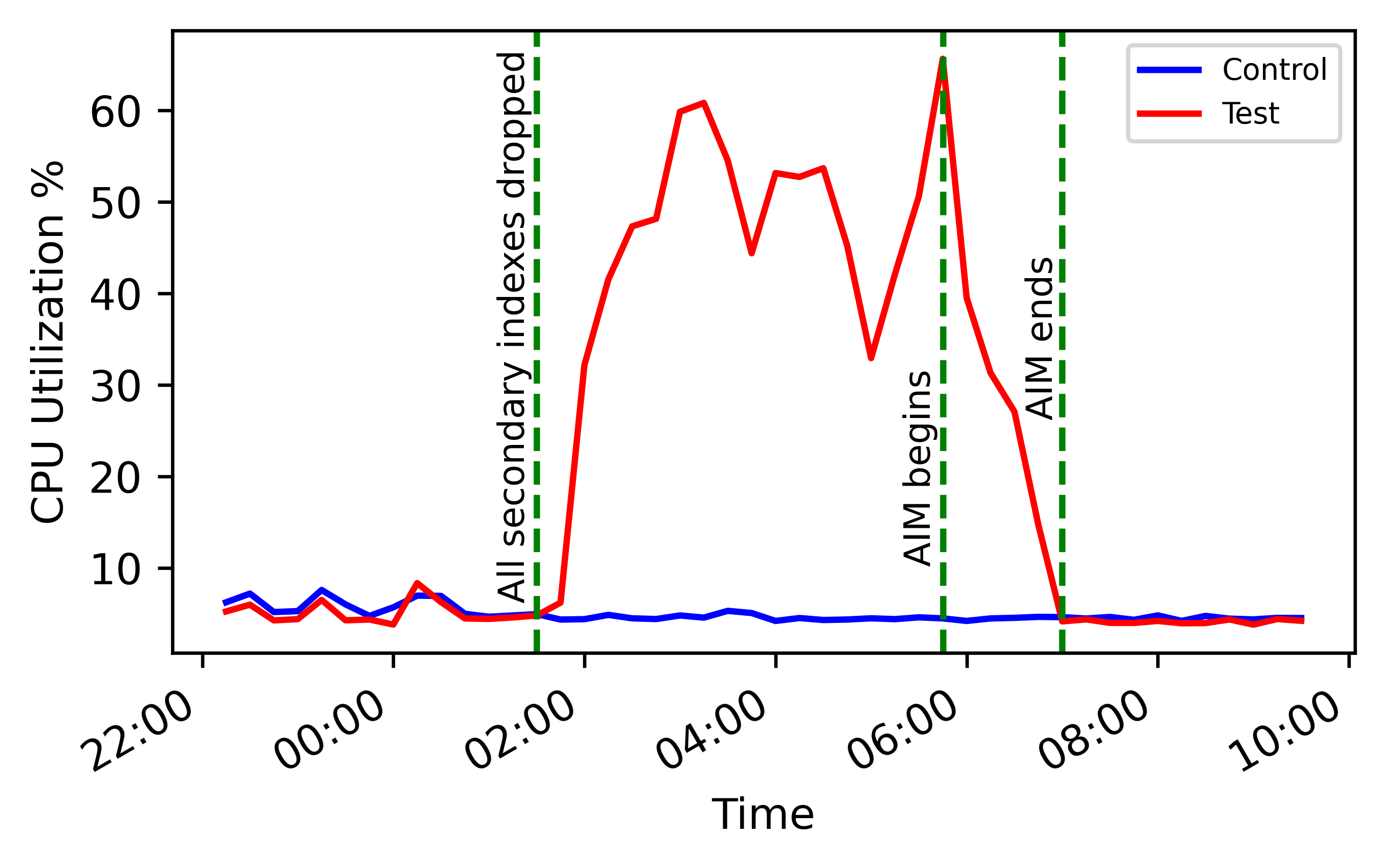}
        \subcaption{CPU utilization graph for Product A}
        \label{fig:gph1}
    \end{subfigure}
    \hfill
    \begin{subfigure}[b]{0.325\textwidth}
        \includegraphics[width=\textwidth]{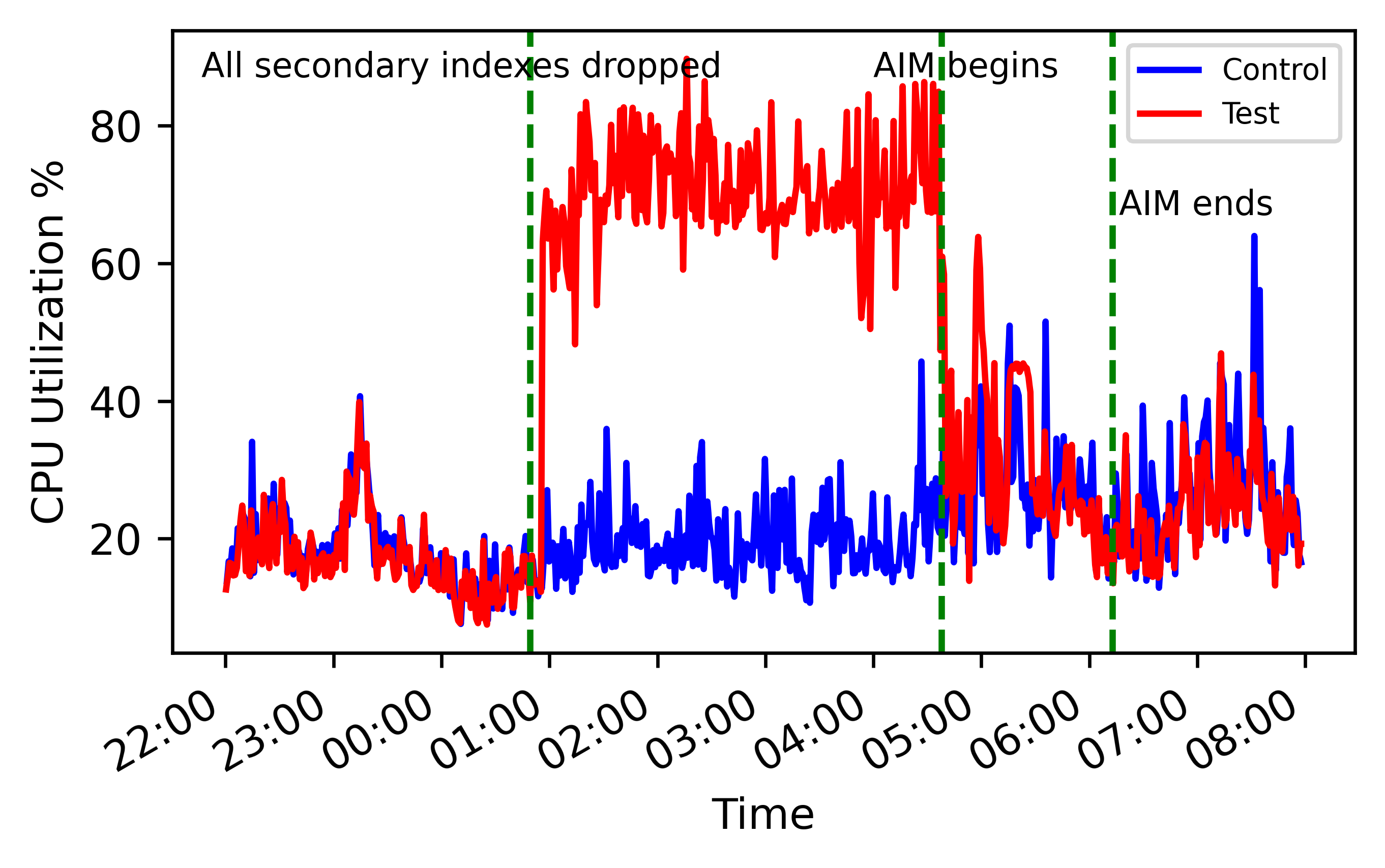}
        \subcaption{CPU utilization graph for Product B}
        \label{fig:gph6}
    \end{subfigure}
    \hfill
    \begin{subfigure}[b]{0.325\textwidth}
        \includegraphics[width=\textwidth]{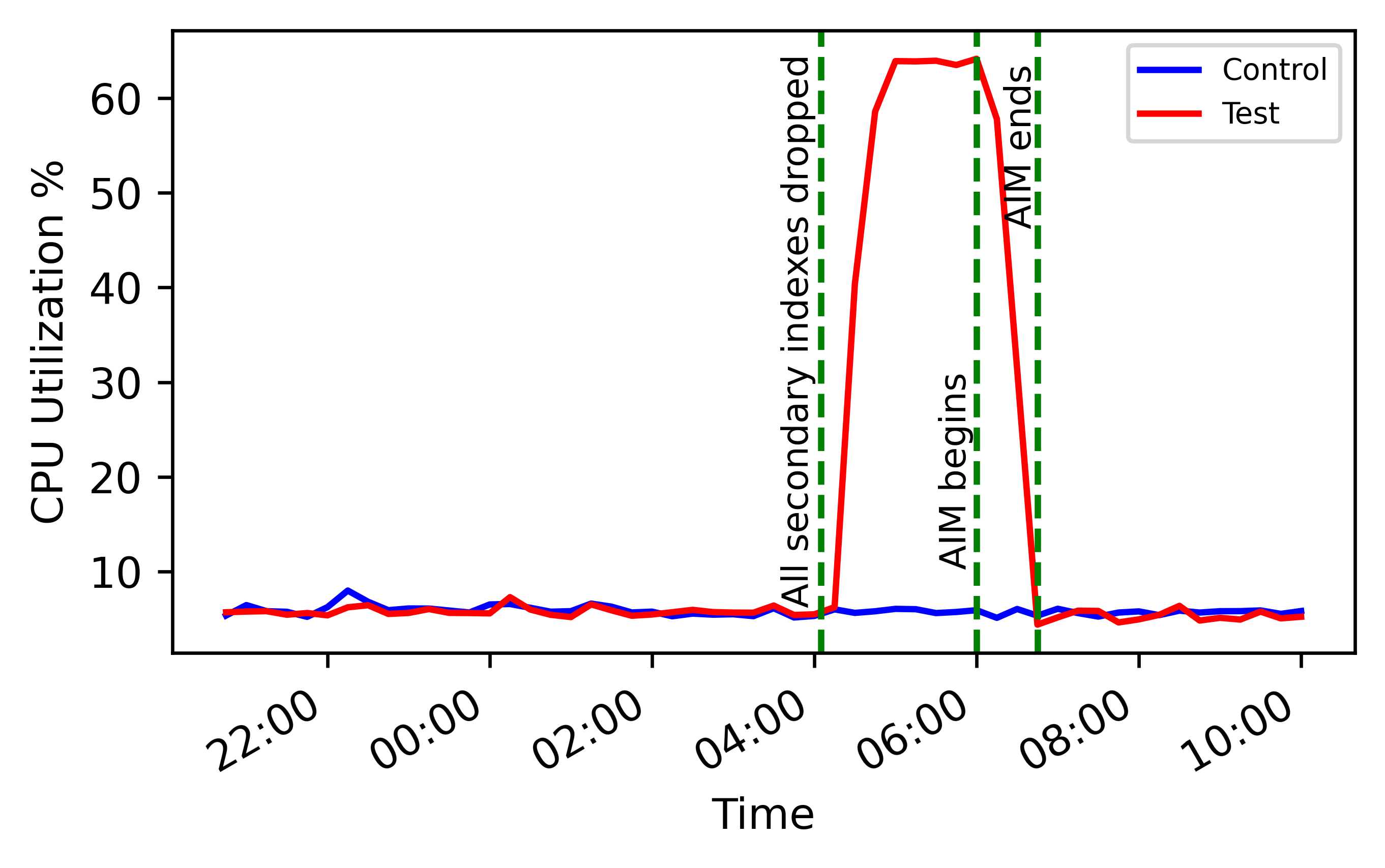}
        \subcaption{CPU utilization graph for Product C}
        \label{fig:gph8}
    \end{subfigure}
    \hfill
    \begin{subfigure}[b]{0.325\textwidth}
        \includegraphics[width=\textwidth]{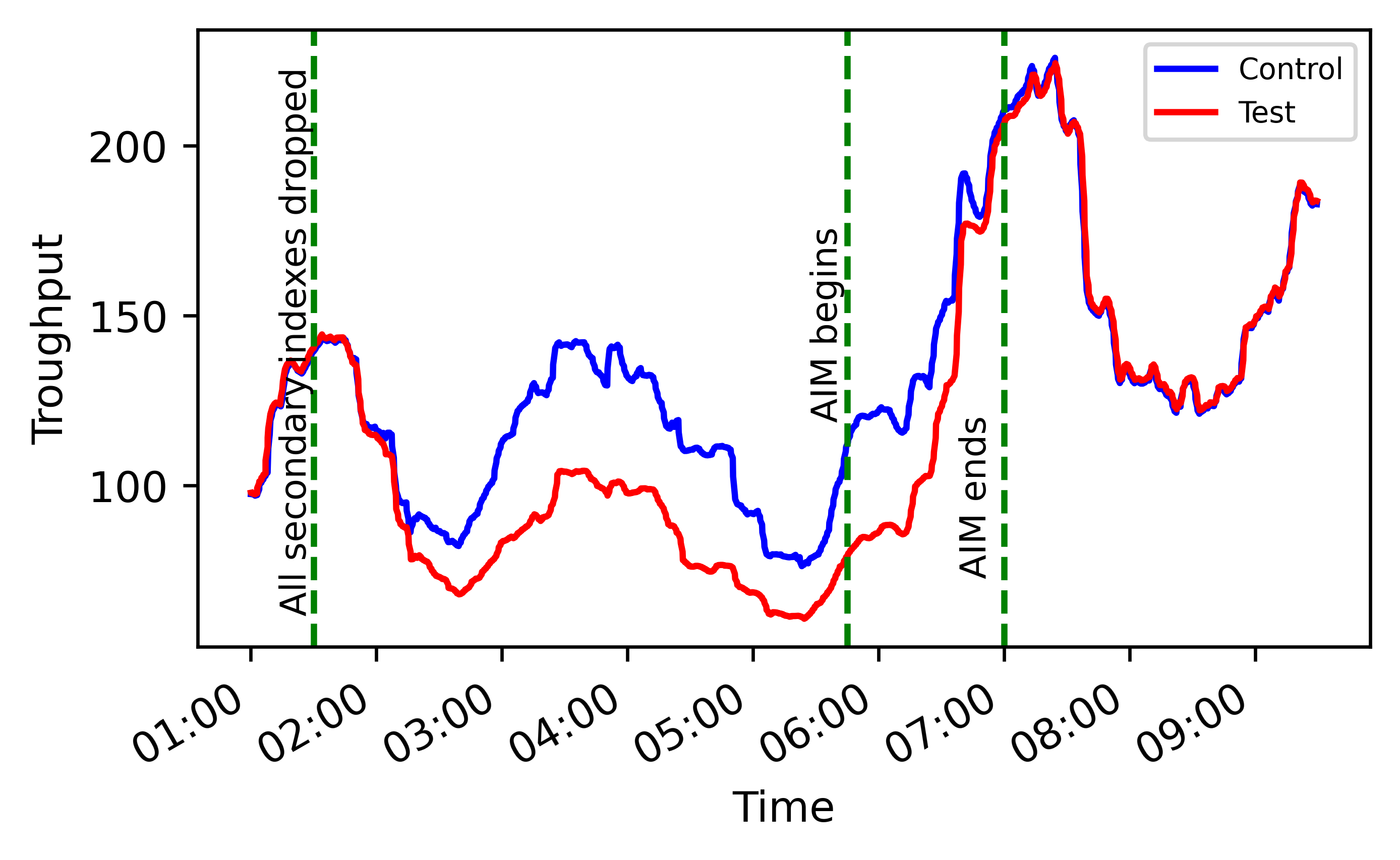}
        \subcaption{Throughput graph for Product A}
        \label{fig:gph3}
    \end{subfigure}
    \hfill
    \begin{subfigure}[b]{0.325\textwidth}
        \includegraphics[width=\textwidth]{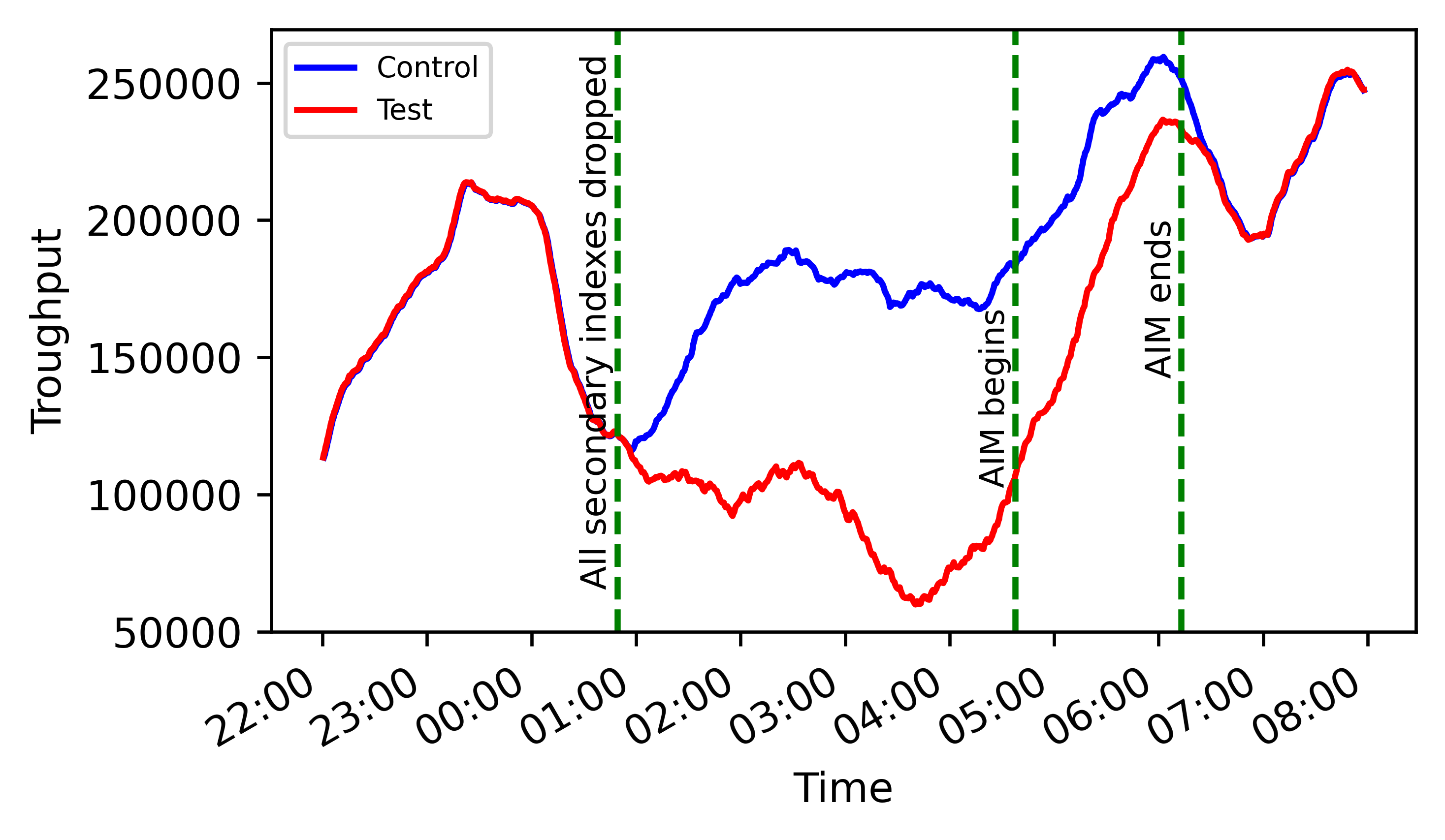}
        \subcaption{Throughput graph for Product B}
        \label{fig:gph7}
    \end{subfigure}
    \hfill
    \begin{subfigure}[b]{0.325\textwidth}
        \includegraphics[width=\textwidth]{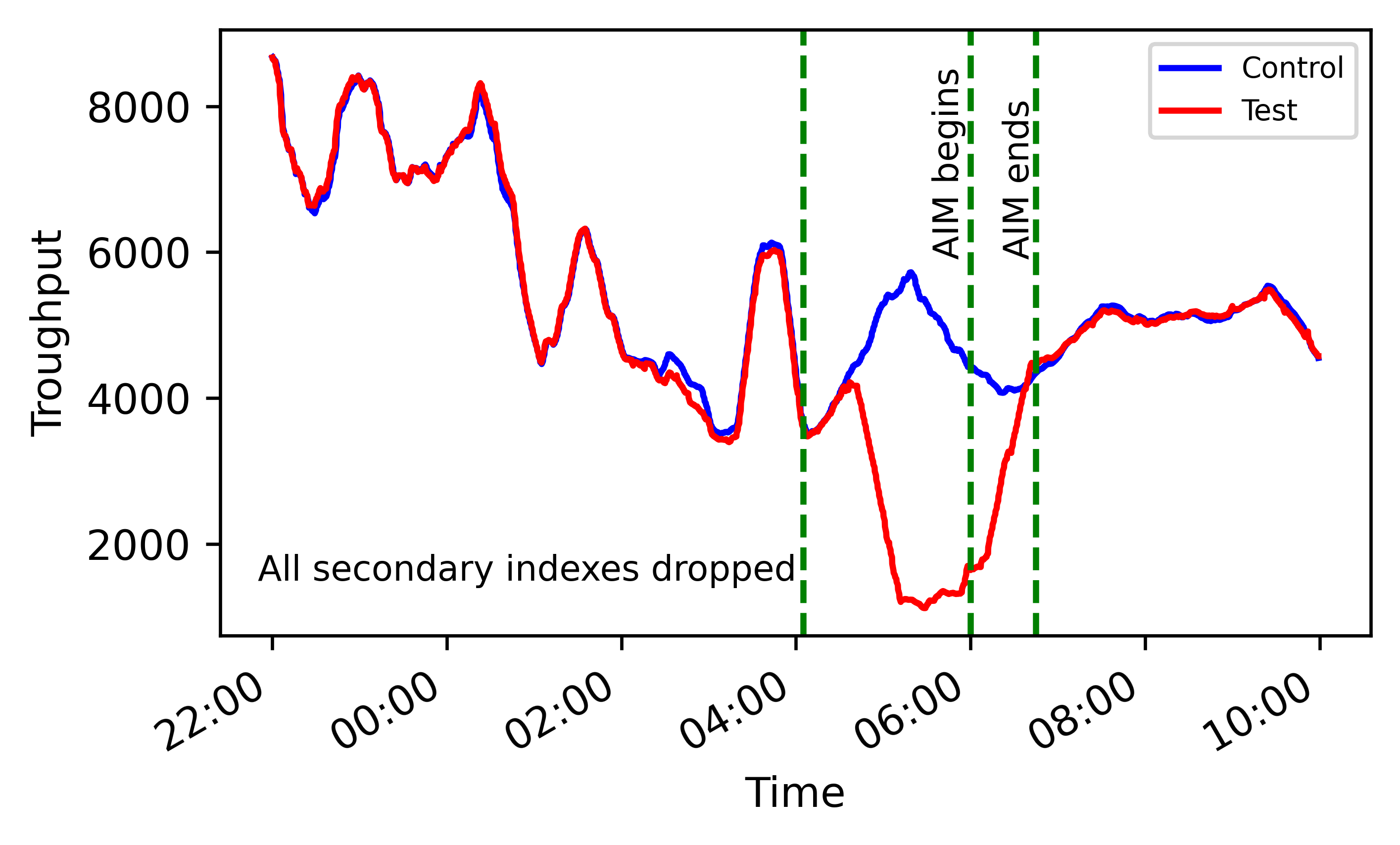}
        \subcaption{Throughput graph for Product C}
        \label{fig:gph9}
    \end{subfigure}
    \caption{CPU utilization \& throughput profiles before and after AIM execution}
    \label{fig:gph10}
\end{figure*}

AIM has been deployed to transactional workloads served by MySQL at Meta which powers a variety of real-time applications. It supports both storage engines; InnoDB (B+ trees) and RocksDB (LSM trees). The workload generated by these applications is diverse and most databases receive dynamic ad-hoc updates. In order to compare AIM's performance against manual tuning, we conducted several tests on representative production workloads where all secondary indexes were removed and AIM was allowed to add them from scratch after analyzing the workload. Not only did AIM achieve performance comparable to manual tuning, it did so using fewer indexes in most cases. Metadata about these representative workloads is provided in \autoref{tab:perftab}. The Jaccard similarity index between the sets of indexes created by DBAs and AIM is also provided.
\begin{table*}[htbp]
  \caption{Performance comparison between DBAs and AIM on production workloads}
  \label{tab:perftab}
  \centering
  \begin{tabular}{*9c}
    \toprule
    \multirow{2}{*}{Product} & \multirow{2}{*}{Tables} & \multirow{2}{*}{Join Queries} & \multirow{2}{*}{Workload Type} & \multicolumn{2}{c}{Index Count} & \multicolumn{2}{c}{Total Indexes Size} & \multirow{2}{*}{Jaccard Similarity} \\
    \cmidrule(lr){5-6} \cmidrule(lr){7-8}
    & & & & DBA & AIM & DBA & AIM \\
    \midrule
    Product A & 147 & 67 & Write Heavy & 248 & 217 & 149.76 GiB & 119.49 GiB & 0.72 \\
    Product B & 184 & 733 & Read Heavy & 287 & 291 & 7.69 GiB & 9.14 GiB & 0.81\\
    Product C & 42 & 25 & Balanced & 51  & 49 & 26.23 GiB & 23.11 GiB & 0.89\\
    Product D & 16 & 18 & Write Heavy & 56 & 52 & 10.27 GiB & 8.42 GiB & 0.61\\
    Product E & 51 & 41 & Read Heavy & 109 & 82 & 688.16 GiB & 517.01 GiB & 0.68\\
    Product F & 5 & 10 & Read Heavy & 33 & 34 & 193.83 MiB & 102.52 MiB & 0.97\\
    Product G & 79 & 386 & Balanced & 232 & 220 & 868.64 GiB & 815.33 GiB & 0.76\\
    \bottomrule
  \end{tabular}
\end{table*}
AIM was able to achieve performance at par with manually tuned databases, even for highly optimized databases with dedicated DBAs. This performance is demonstrated in \autoref{fig:gph10} for Product A, B \& C listed in \autoref{tab:perftab}. Each graph shows the CPU utilization on the machines and the observed throughput as we drop all secondary indexes and initiate AIM which recreates them from scratch on the test setup (shown in red). The control and test setups are hosted on separate machines with the exact same hardware capabilities, data and workload. The control setup (shown in blue) is left untouched with indexes created manually by DBAs.

\subsection{Comparison with other state-of-the-art algorithms}
\label{subsec:sota}
We also benchmarked AIM against the state-of-the-art algorithms from industry (DTA \cite{dta}) and academia (Extend \cite{schlosser2019efficient}) using the open-source framework developed by Kossmann et al \cite{kossmann2020magic} on PostgreSQL v12.11. The framework supports eight algorithms (AutoAdmin \cite{chaudhuri1997efficient}, CoPhy \cite{dash2011cophy}, DB2Advis \cite{valentin2000db2}, DTA \cite{dta}, Dexter \cite{dexter}, Drop \cite{whang1987index}, Extend \cite{schlosser2019efficient} and Relaxation \cite{bruno2005automatic}) but we chose only two to compare against for clarity in our graphs. DTA and Extend were reported as the best performing algorithms in industry and academia, respectively. The framework utilizes HypoPG \cite{hypopg} built for PostgreSQL and the optimizer does not account for index maintenance costs which is why purely analytical benchmarks are chosen to compare the index selection prowess of different algorithms.

\begin{figure*}[htbp]
    \centering
    \begin{minipage}{0.65\textwidth}
        \centering
        \subcaptionbox{TPC-H (SF 10): Solution quality. Cost relative to processing without indexes.\label{fig:tpch1}}
            {\includegraphics[width=0.494\textwidth]{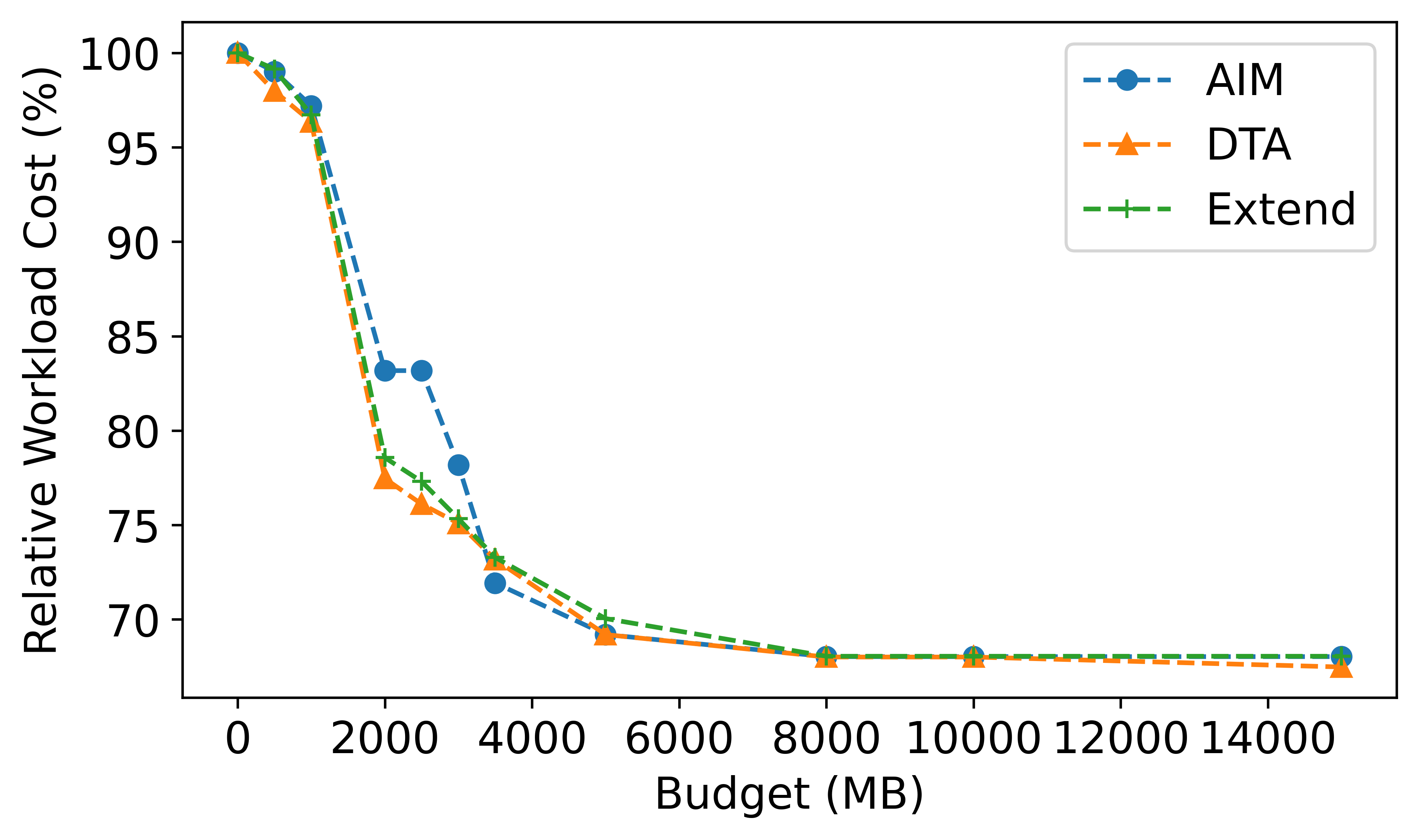}}
        \subcaptionbox{TPC-H (SF 10): Observed runtime.\label{fig:tpch2}}
            {\includegraphics[width=0.494\textwidth]{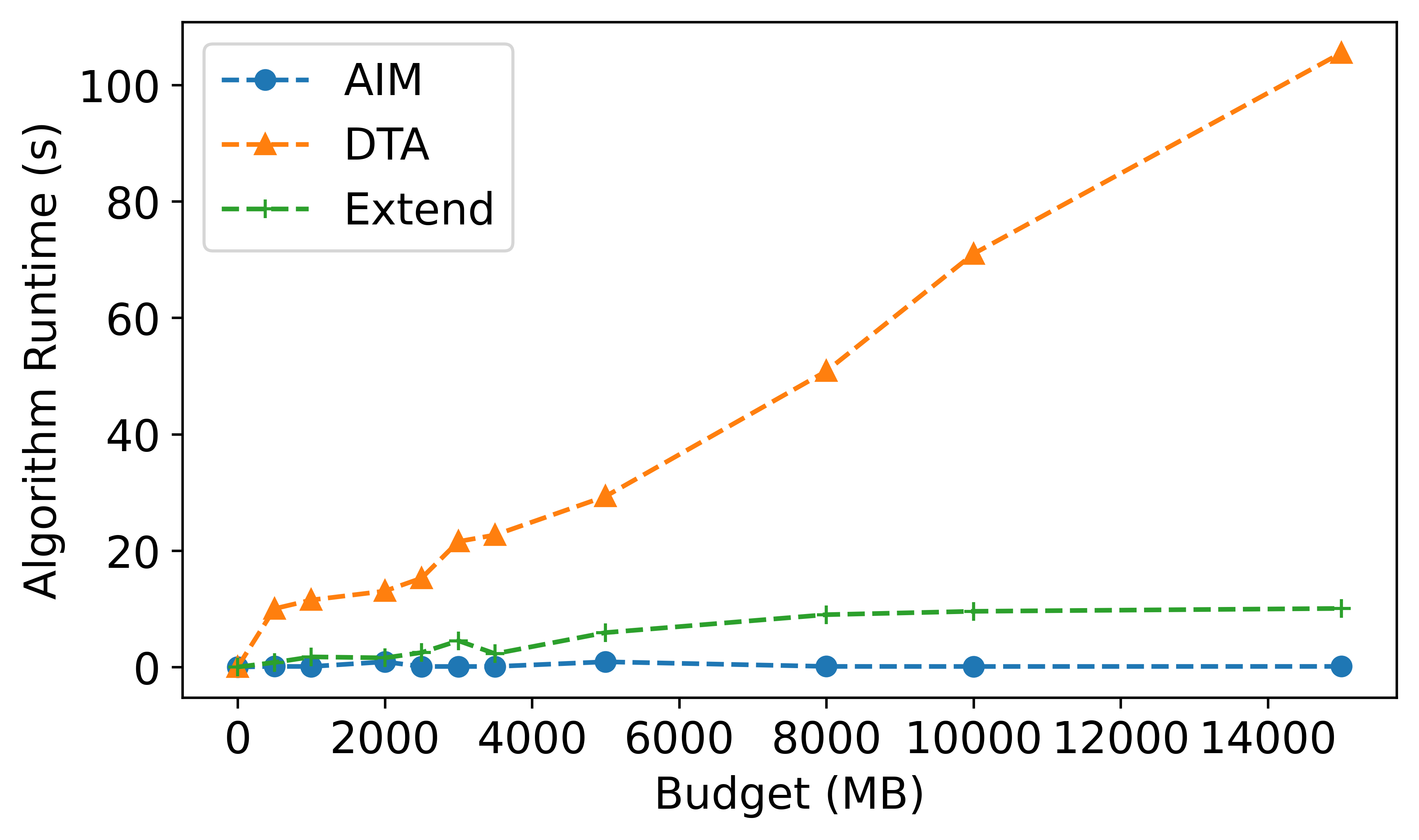}}
        \subcaptionbox{JOB: Solution quality. Cost relative to processing without indexes.\label{fig:job1}}
            {\includegraphics[width=0.494\textwidth]{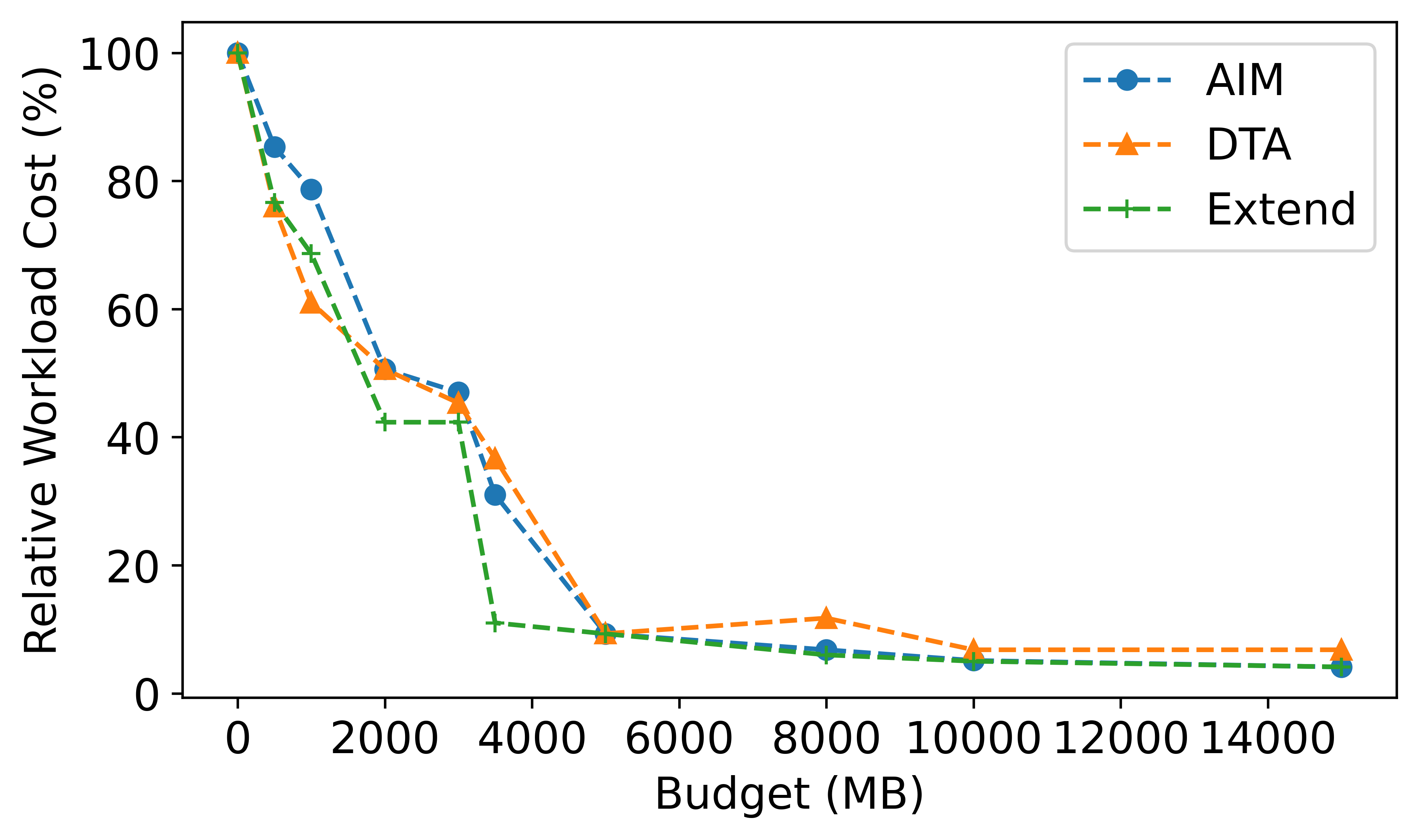}}
        \subcaptionbox{JOB: Observed runtime.\label{fig:job2}}
            {\includegraphics[width=0.494\textwidth]{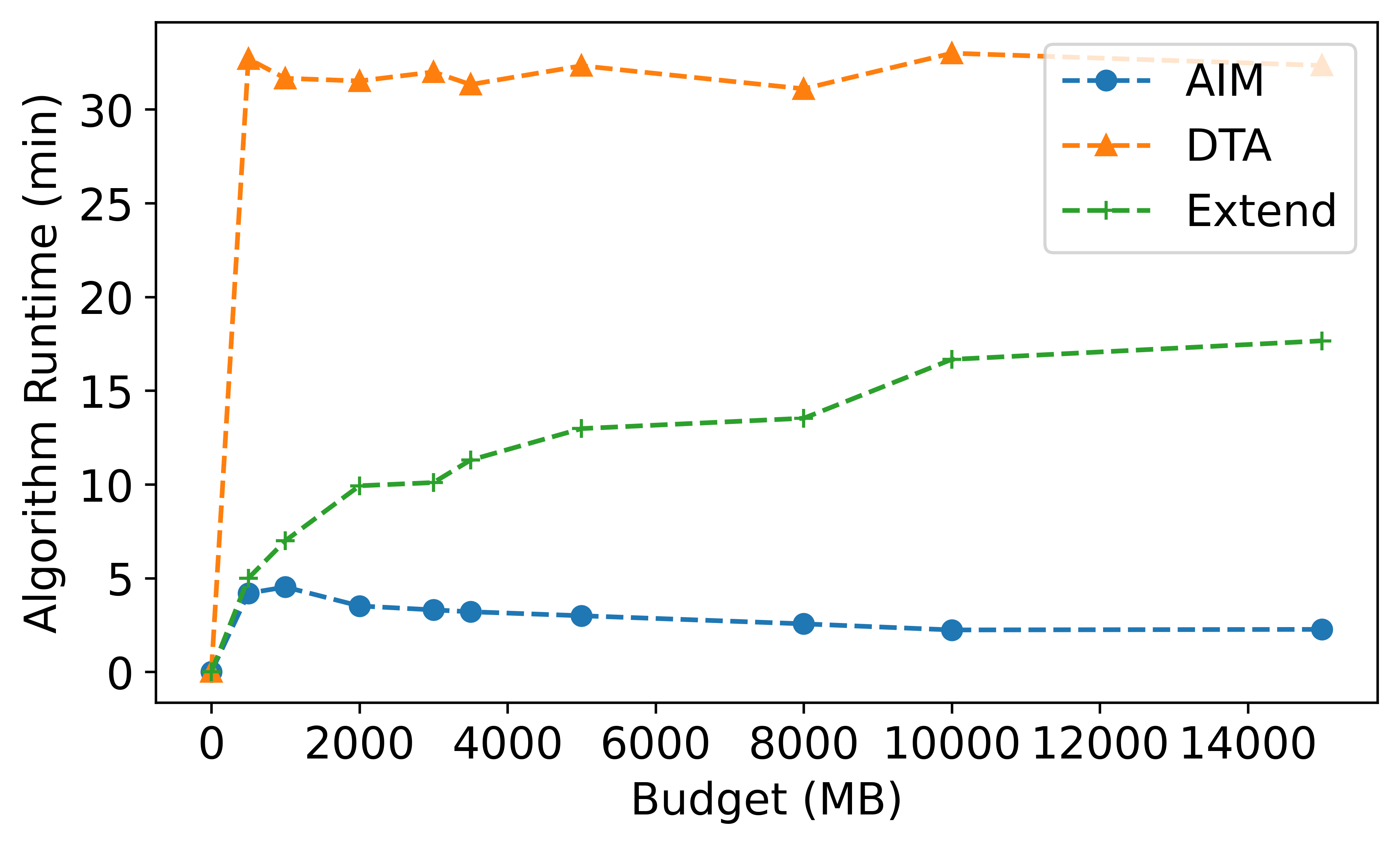}}
        \caption{Estimated workload processing costs \& algorithm runtimes}
        \label{fig:benchperf}
    \end{minipage}
    \begin{minipage}{0.32\textwidth}
        \centering
        \subcaptionbox{Only those queries are shown where indexes had an effect.\label{fig:tpchq1}}
            {\includegraphics[width=0.933\textwidth]{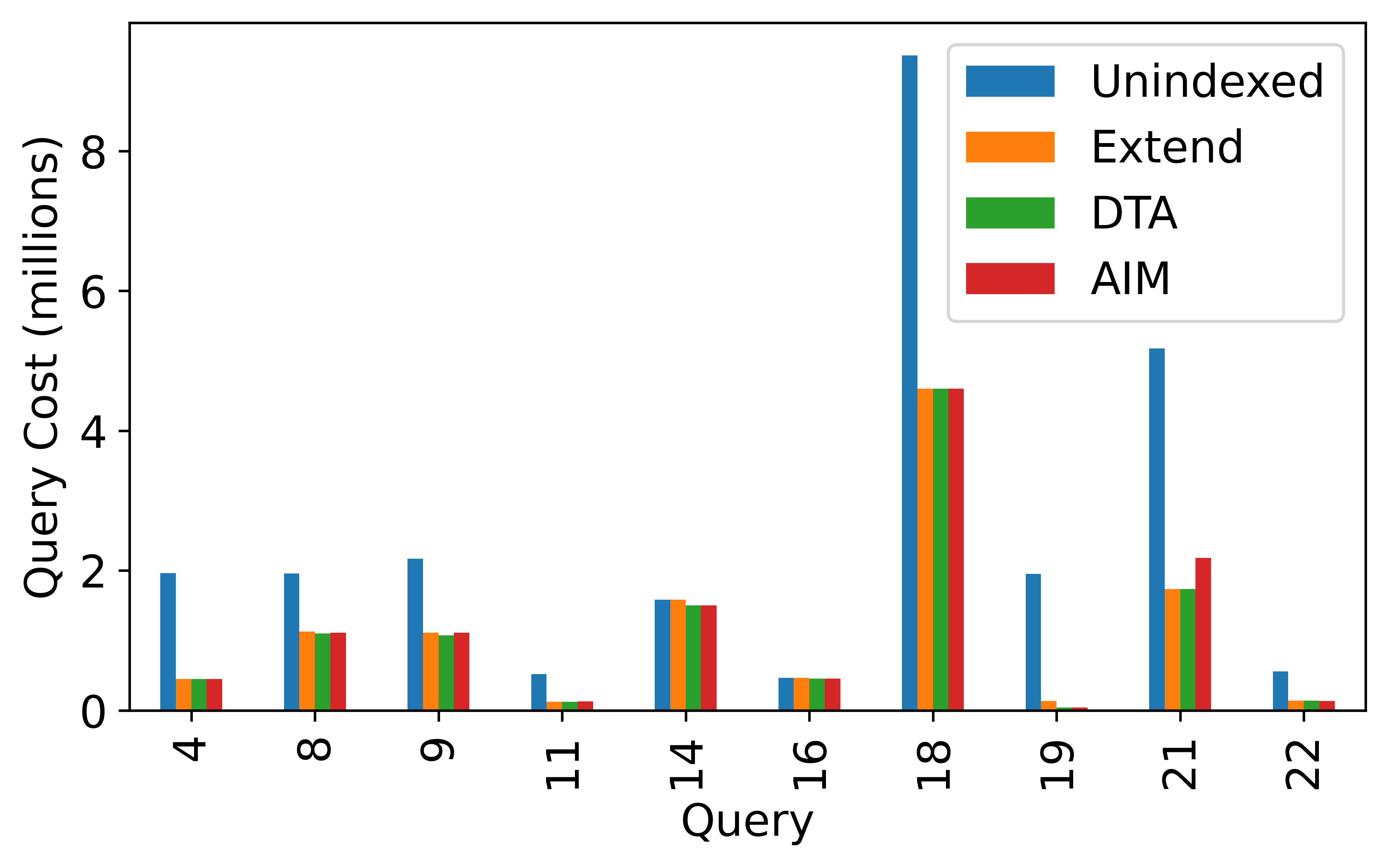}}
        \subcaptionbox{Only expensive queries are shown in log scale.\label{fig:tpchq2}}
            {\includegraphics[width=0.933\textwidth]{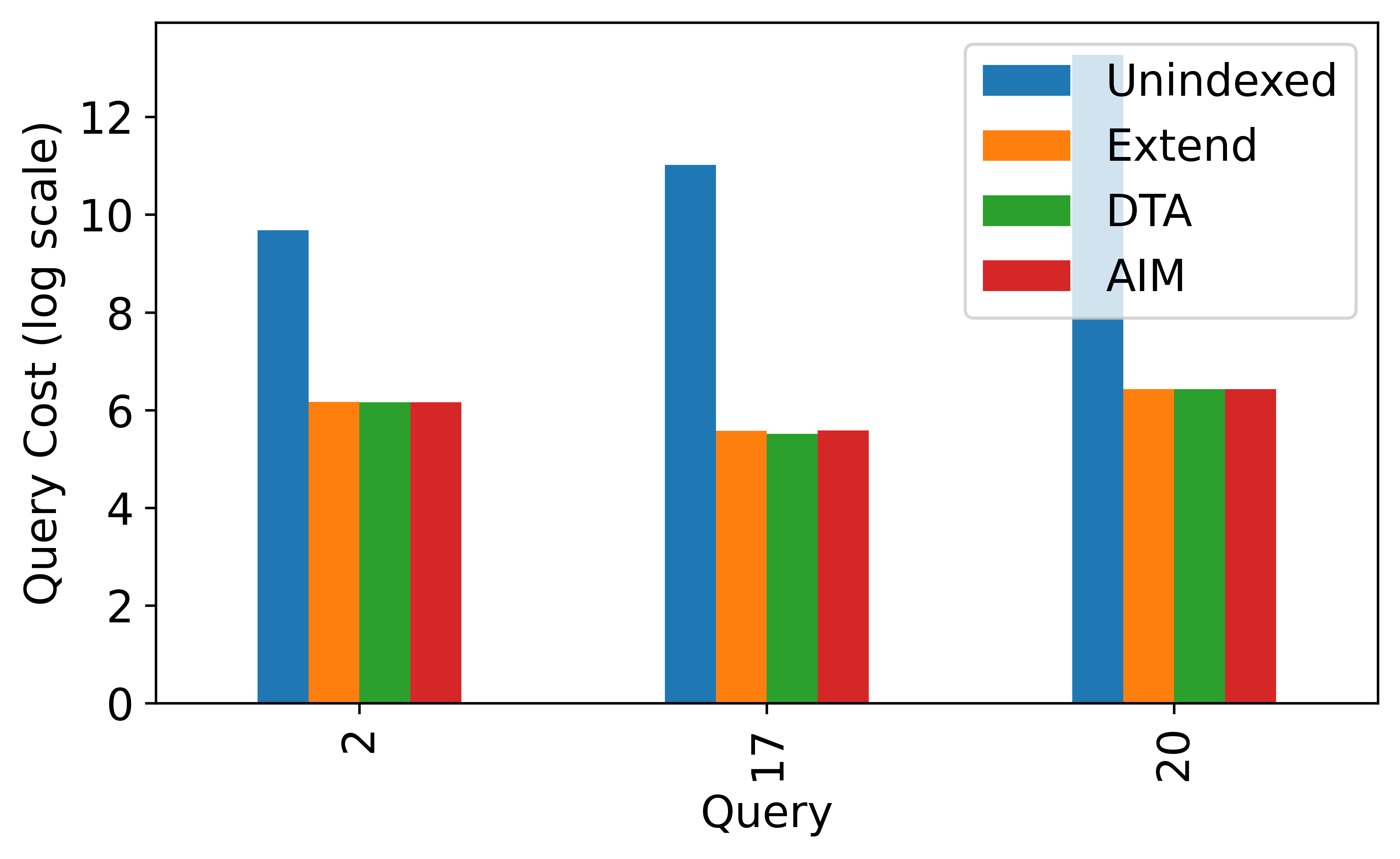}}
        \caption{Query processing costs for TPC-H (scale factor 10) on PostgreSQL.}
    \end{minipage}
\end{figure*}

The main challenge which hindered a proper comparison with DTA was its evaluation strategy, which became prohibitively expensive \cite{kossmann2020magic} when considering indexes of width $>$ 3 for complex workloads (TPC-DS and JOB). Therefore, we limited the setup to only consider indexes of maximum width 4 for TPC-H and width 3 for JOB. This experiment is presented in ~\autoref{fig:benchperf} where the optimizer \emph{estimated} cost of processing a workload relative to unindexed processing cost (y-axis) is plotted against different values of storage budgets (x-axis) for the TPC-H and JOB benchmarks in ~\autoref{fig:tpch1} \& ~\ref{fig:job1}. The algorithm runtimes (y-axis) are also plotted against storage budgets (x-axis) in ~\autoref{fig:tpch2} \& ~\ref{fig:job2}. Graphs from TPC-DS benchmark followed the same trend and are not included to save space.

The compromise implemented by AIM is evident from the benchmark results. It trades solution granularity for faster convergence. For lower storage budgets, AIM's solution is sometimes not as good as DTA or Extend since it doesn't explore the search space in a more granular fashion. However, as soon as the budget constraints are reasonably relaxed, it's solution quality is better or at par with the state-of-the-art algorithms. The main thing to notice here is the \emph{relatively cheap and stable runtime} offered by AIM since it can quickly detect the most efficient indexes for a query by utilizing its structural information. Most state-of-the-art algorithms explore configurations at a fine granularity; often greedily adding one column at a time. This leads to prohibitively large runtimes and missing out on optimal solutions where intermediate configurations are rejected for lack of incremental gain.

~\autoref{fig:tpchq1} and ~\autoref{fig:tpchq2} show the performance of the solutions chosen by different algorithms across individual queries in the TPC-H benchmark (scale factor 10) with a budget constraint of 15 GB. Relatively expensive query costs are shown in log scale. It can be seen that the performance for individual queries is pretty similar across all algorithms, except for Q21. AIM chose a covering index for Q21 and PostgreSQL query optimizer assigned it a higher estimated cost. However, the actual query execution costs were similar.

\subsection{Impact of the join parameter}
Another interesting aspect of the AIM algorithm is its treatment of join queries. The only way to ensure that the most optimal join order gets picked for each query is to have indexes that support all potentially efficient join orders. On top of that, the optimizer should be able to select the optimal join order in the presence of these indexes in an efficient manner. Neither of these prerequisites can be fulfilled in practice due to various constraints \cite{innolim, leis2015good}. Therefore, AIM uses the join parameter $j$ and limits the number of explored configurations as described in ~\autoref{sub:joins}.
\begin{figure}[htbp]
  \centering
  \includegraphics[width=0.52\textwidth]{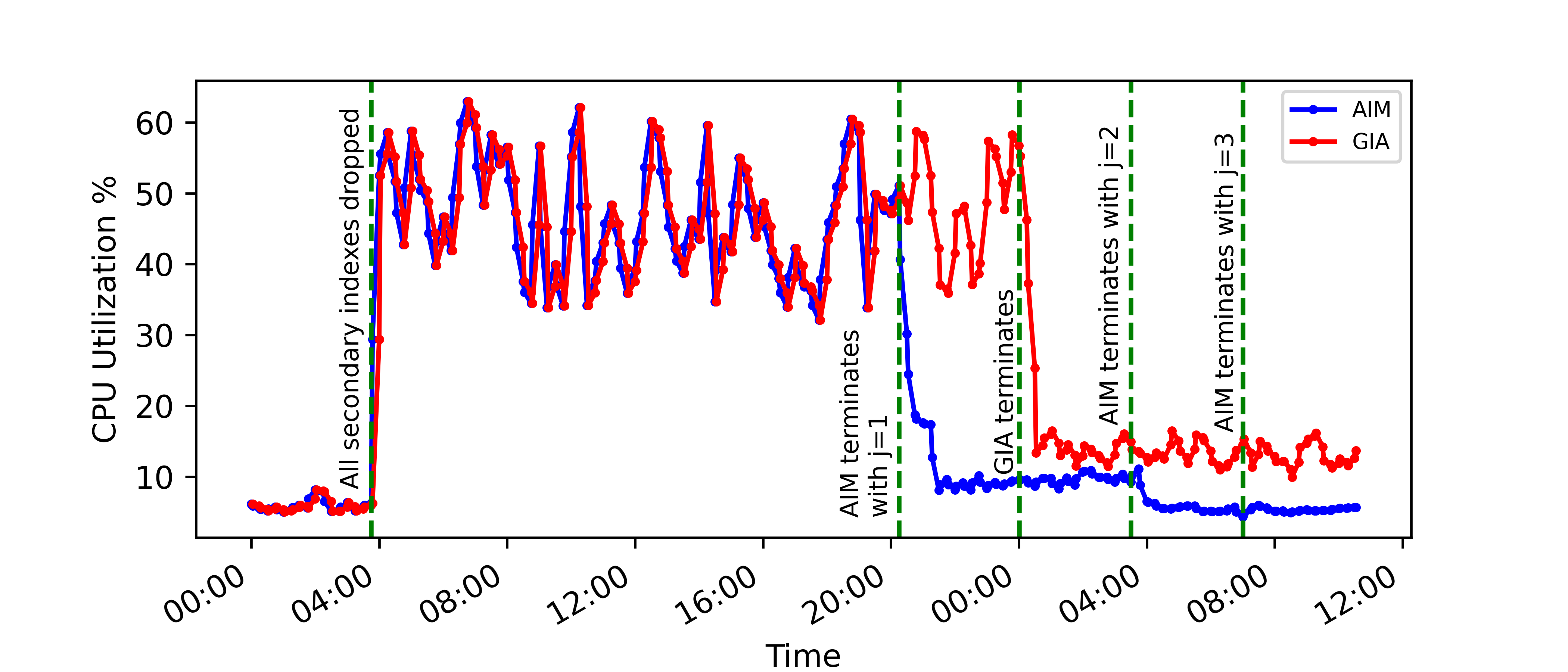}
  \caption{Effect of join parameter}
  \label{fig:gph4}
\end{figure}
In this experiment, all secondary indexes were removed from identical databases on two separate machines. A constant perioidically repeating workload is replayed on both machines. For this experiment, we chose a transactional workload with many complex join queries to demonstrate AIM's impact. AIM progressively creates secondary indexes with increasing values of the join parameter $j$ on one machine and a greedy incremental algorithm (Extend) creates them on the other. The difference in performance between the two solutions is visible from ~\autoref{fig:gph4}. AIM's solution achieves 27\% better throughput compared to the greedy algorithm (labeled GIA) and results in 4.8\% lower CPU utilization. This is because the quantum of configuration exploration is limited to one column at every step and the incremental benefit might not be justified by the cost function. Consider a join clause between a pair of tables with three sub-predicates. It is possible that any combination of two sub-predicates is not selective enough but a combination of all three is highly selective. In such cases, greedy algorithms would not be able to detect efficient join orders for the query. Some machine learning algorithms \cite{zhou2022autoindex} try to explore suboptimal intermediate configurations in a probabilistic fashion but utilizing the query structures for identifying changes to the existing configuration yields definitive results.

Another important aspect demonstrated by the experiment is the change in performance with respect to increasing values of $j$. The solution with $j = 2$ results in 16\% better throughput than the one AIM arrived at with $j = 1$. The gain from increasing $j$ from 2 to 3 was insignificant. In theory, we can construct queries on top of synthetic data distributions that would require $j > 3$ to achieve optimal performance but we haven't seen such cases in production so far.

Please note that indexes were created incrementally with sleeps in between in order to clearly observe the impact on the graphs presented in ~\autoref{fig:gph10} \& ~\autoref{fig:gph4}.

\subsection{Continuous index tuning}
AIM is not only good at bootstrapping secondary index creation, it also does particularly well at detecting indexes when the workload changes significantly. Most of the times, expensive queries result from new code pushes where developers forget to create supporting secondary indexes beforehand. It is evident from ~\autoref{fig:benchperf} that AIM's runtime is significantly lower compared to other state-of-the-art algorithms. Therefore, we naïvely achieve continuous tuning by running AIM periodically. Continuous tuning has resulted in savings of approximately 2\% CPU capacity required for serving OLTP workloads with roughly 31\% of the improved queries resulting in at least an order of magnitude improvement. Adopting features like controllable overhead and phased index profiling from advanced continuous tuning algorithms like Colt \cite{schnaitter2007line} will be addressed by future work. 

\section{Supporting Components}
\label{section:support_comp}
\subsection{Continuous statistics export}
The fault tolerant MySQL offering at Meta \cite{yadavflexiraft} replicates databases across multiple machines. Read queries are served by one of the many available replicas. Therefore, statistics need to be gathered from all of these machines and aggregated to get a holistic view of the workload for a given database. This is achieved by a daemon process which periodically queries every single machine running on the fleet and exports it to Meta’s internal data warehouse. The export is achieved by a pub-sub system (similar to Apache Kafka \cite{thein2014apache}) and complex analytics can be run almost instantaneously.

\subsection{MyShadow}
\label{section:clone}
MyShadow framework is a test environment provider which creates a temporary logical copy of the database (clone) and can replay traffic from production database instances onto test instances. This framework plays an important role in catching significant regressions that are only possible to detect in a production-like environment. It helps us make sure that there are no side-effects of running AIM on sensitive production workloads. The MyShadow framework was used to setup the experiments presented in ~\autoref{section:expt}. It has the ability to sample the data and workload being replayed. Therefore, it is suitable for providing economical test beds.

\subsection{Continuous Regression Detector}
The continuous regression detector is an independent process that looks for regressions in performance of queries across the fleet due to query optimizer plan changes. It relies on the average CPU time consumption by a normalized query over a period of time. If a regression is detected due to an index added by automation, it is flagged for removal.

\section{Operational Experience and Lessons}
\label{section:lessons}
\paragraph{Query optimizer nuances}
Our work on AIM gave us a lot of insight into the tradeoffs that database engines make when executing queries. The work of Chaudhuri and Narasayya \cite{chaudhuri1997efficient} pioneered the concept of consulting the query optimizer for index selection. It was a huge step forward but optimizers (even in mature databases) make lots of mistakes \cite{wu2013predicting}. There are three interesting aspects to consider. First, the optimizer might pick a sub-optimal plan which can prevent usage of beneficial indexes. We have seen several cases in production where AIM's index suggestions which offered significant reduction in query processing cost were rejected by MySQL's query optimizer. Second, optimizer switches are often used to influence the query optimizer plan selection. The number of candidates generated can be significantly reduced if candidate generation takes the value of these switches into account. Features like index skip scan \cite{skipscan}, index merge intersections \cite{imo} etc. maybe switched off for a subset of databases due to correctness and performance bugs \cite{skipscanbug,imobug}. Making the index candidate generation aware of their values improves the efficiency of the algorithm. Third, most modern algorithms do not scale well with increasingly complex workloads as a huge fraction of the runtime is spent making optimizer calls \cite{dash2011cophy, papadomanolakis2007efficient}. In fact, while benchmarking AIM against state-of-the-art algorithms (~\autoref{subsec:sota}), we had to set a really high timeout for DTA when exploring candidates of width greater than or equal to 3 for TPC-DS and JOB benchmarks.

\paragraph{Economics of sharding}
Databases that cannot be served by the resources on a single machine are partitioned horizontally. Each constituent partition which consists of only a subset of the data is called a shard. Since our deployment mandates common physical design across all shards of a database, the economics of the index selection problem were adjusted for heavily sharded databases. This is because the improved queries might not frequently execute on all consituent shards but each shard has to pay for the storage and maintenance of all indexes. Similarly, query regressions which occur on a subset of the shards are difficult to detect in real-time without comprehensive validation across all shards. We provide configurable parameters for performance sensitive databases to perform comprehensive validation and rely on the continuous regression detection to revert unwanted changes to the physical design.

\paragraph{Compute placement and regression management}
The AIM process does not run on individual database hosts and a centralized coordinator kicks off the tuning process for a database if it detects inefficient queries. This setup prevents compute wastage on the entire fleet. The continuous regression detector is also an off-host centralized process that serves as an indispensable protection mechanism because some portions of the workload may repeat after a very long duration (e.g. monthly report generation). Therefore, reverting any regressions quickly is of paramount importance and algorithms which take hours to converge cannot be relied upon.

\section{Related \& Future Work}
\label{section:relatedw}

The problem of automatically adding secondary indexes has been studied since 1970 \cite{palermo1970quantitative}. A variety of algorithm classes have been developed over time with varying degrees of production readiness. Based on the methodology used for candidate selection, these algorithms can be broadly classified into two classes; imperative and declarative \cite{kossmann2020magic}. The imperative class of algorithms either start with an empty configuration and iteratively add indexes or start with a very large configuration which is then iteratively reduced. Most of these algorithms rely heavily on usage of “what-if” indexes \cite{chaudhuri1998autoadmin}. Papadomanolakis et al. show that index selection algorithms spend 90\% of their runtime in the optimizer, on average \cite{papadomanolakis2007efficient}. There is also a significant body of work that tries to reduce the number of “what-if” calls to the optimizer \cite{papadomanolakis2007efficient,bruno2011configuration,wu2022budget}. The problem is complicated by the fact that optimizer decisions are sometimes unreliable \cite{leis2015good,wu2013predicting} and can lead to severe plan regressions \cite{borovica2012automated,das2019automatically}.

The occasional shortcomings of the optimizer brings the focus to the declarative class of algorithms which either employ machine learning or linear programming. The linear programming models \cite{dash2011cophy,caprara1995exact,caprara1996branch} solve optimization problems by specifying the optimization goal and constraints as linear equations. These formulations do not scale well with the size of the problem \cite{schlosser2019efficient} and often restrict the solution space (e.g. restricting index width). 

More recently, machine learning algorithms have been utilized for index management \cite{basu2016regularized, sharma2018case, sadri2020online, perera2021dba, wang2021udo, wu2022budget, ding2019ai}. There are two common themes across these algorithms. They either use deep reinforcement learning (DRL) \cite{sharma2018case, perera2021dba, wang2021udo, wu2022budget} for the index selection problem or craft a deep learning (DL) model \cite{hasan2020deep, ding2019ai} to improve query cost prediction as compared to the optimizer. 

The problem with DRL algorithms is that they are computationally expensive \cite{zhou2022autoindex}. They usually rely on a lot of training data and are not well suited for dynamic workloads due to their higher runtimes. For e.g., the implementation of Sharma et al. \cite{sharma2018case} required 8 hours of training on CPUs for a small workload like TPC-H \cite{kossmann2020magic}. The DL models on the other hand operate at the query level and are better suited for analytical queries. Most of these methods do not model the index maintenance overheads. However, there are many useful ideas that can be borrowed from machine learning techniques to AIM, such as training a classifier to predict comparative performance of execution plans rather than relying on the optimizer as described by Ding et al \cite{ding2019ai}. In fact, the technique described by UDO \cite{wang2021udo} can be used to build environment specific benchmarks based on actual query performance since it generates \emph{guaranteed} optimal plans via an expensive process. Furthermore, several avenues to improve cost models of open source query optimizers based on the utilization of interesting sort orders \cite{guravannavar2012sort} have been identified. 

AIM’s candidate index selection falls under the imperative class but it tries to limit its reliance on the query optimizer by utilizing the structural properties of queries. AIM compromises on the granularity of the solution in favour of converging really fast. Since the importance of a query is a subjective attribute, it is not possible to ascertain it automatically. Therefore, our approach does not try to explore configurations that are suboptimal for constituent normalized queries but optimal for the overall performance. To the best of our knowledge, relaxation \cite{bruno2005automatic} is the only other modern algorithm which utilizes the query structure to a significant extent but its approach of iteratively pruning indexes after starting from a very large candidate set gives it a prohibitively expensive runtime. Unlike Relaxation, AIM does \emph{not} need any extra instrumentation to the query optimizer.

\section{Concluding remarks}
In this paper, we presented AIM which is an autonomous index management solution for SQL databases. It works well for transactional workloads in production and can scale with increasingly complex workloads. A detailed discussion of the AIM algorithm proposes a novel way of exploring the search space of candidate indexes and presents a systematic treatment of complex join queries for the first time. The compromise between fast convergence and reduced solution exploration granularity has been explained thoroughly. The challenges and lessons learned from the production deployment of AIM have been shared with the hope that engineers looking to automate index management on their SQL databases can utilize this information in their implementations. Experiments described in the paper compare AIM against DBAs and other modern algorithms.

\bibliographystyle{ieeetr}
\bibliography{sample-base}

\begin{thebibliography}{10}

\bibitem{agrawal2005database}
S.~Agrawal, S.~Chaudhuri, L.~Kollar, A.~Marathe, V.~Narasayya, and M.~Syamala,
  ``{Database tuning advisor for Microsoft SQL Server 2005},'' in {\em
  Proceedings of the 2005 ACM SIGMOD International Conference on Management of
  Data}, pp.~930--932, 2005.

\bibitem{bruno2007online}
N.~Bruno and S.~Chaudhuri, ``An online approach to physical design tuning,'' in
  {\em 2007 IEEE 23rd International Conference on Data Engineering (ICDE)},
  pp.~826--835, IEEE, 2007.

\bibitem{chaudhuri1997efficient}
S.~Chaudhuri and V.~R. Narasayya, ``{An Efficient Cost-Driven Index Selection
  Tool for Microsoft SQL Server},'' in {\em Proceedings of the 23rd
  International Conference on Very Large Databases (VLDB)}, p.~146–155, 1997.

\bibitem{chaudhuri1998autoadmin}
S.~Chaudhuri and V.~Narasayya, ``{AutoAdmin “What-If” Index Analysis
  Utility},'' in {\em Proceedings of the 1998 ACM SIGMOD International
  Conference on Management of Data}, p.~367–378, 1998.

\bibitem{chaudhuri2007self}
S.~Chaudhuri and V.~Narasayya, ``{Self-Tuning Database Systems: A Decade of
  Progress},'' in {\em Proceedings of the 33rd International Conference on Very
  Large Databases (VLDB)}, pp.~3--14, 2007.

\bibitem{dageville2004automatic}
B.~Dageville, D.~Das, K.~Dias, K.~Yagoub, M.~Zait, and M.~Ziauddin,
  ``{Automatic SQL tuning in Oracle 10g},'' in {\em Proceedings of the 30th
  International Conference on Very Large Databases (VLDB)}, pp.~1098--1109,
  2004.

\bibitem{valentin2000db2}
G.~Valentin, M.~Zuliani, D.~C. Zilio, G.~Lohman, and A.~Skelley, ``Db2 advisor:
  An optimizer smart enough to recommend its own indexes,'' in {\em 2000 IEEE
  16th International Conference on Data Engineering (ICDE)}, pp.~101--110,
  IEEE, 2000.

\bibitem{zilio2004db2}
D.~C. Zilio, J.~Rao, S.~Lightstone, G.~Lohman, A.~Storm, C.~Garcia-Arellano,
  and S.~Fadden, ``{DB2 Design Advisor: Integrated Automatic Physical Database
  Design},'' in {\em Proceedings of the 30th International Conference on Very
  Large Databases (VLDB)}, pp.~1087--1097, 2004.

\bibitem{chaudhuri2004index}
S.~Chaudhuri, M.~Datar, and V.~Narasayya, ``Index selection for databases: A
  hardness study and a principled heuristic solution,'' {\em IEEE Transactions
  on Knowledge and Data Engineering (TKDE)}, vol.~16, no.~11, pp.~1313--1323,
  2004.

\bibitem{leis2015good}
V.~Leis, A.~Gubichev, A.~Mirchev, P.~Boncz, A.~Kemper, and T.~Neumann, ``How
  good are query optimizers, really?,'' {\em Proceedings of the VLDB
  Endowment}, vol.~9, no.~3, pp.~204--215, 2015.

\bibitem{schlosser2019efficient}
R.~Schlosser, J.~Kossmann, and M.~Boissier, ``Efficient scalable
  multi-attribute index selection using recursive strategies,'' in {\em 2019
  IEEE 35th International Conference on Data Engineering (ICDE)},
  pp.~1238--1249, IEEE, 2019.

\bibitem{dta}
S. Chaudhuri and V. Narasayya, {\em Anytime Algorithm of Database Tuning
  Advisor for Microsoft SQL Server}, October 2022.
\newblock
  \url{https://www.microsoft.com/en-us/research/publication/anytime-algorithm-of-database-tuning-advisor-for-microsoft-sql-server/}.

\bibitem{kossmann2020magic}
J.~Kossmann, S.~Halfpap, M.~Jankrift, and R.~Schlosser, ``Magic mirror in my
  hand, which is the best in the land? an experimental evaluation of index
  selection algorithms,'' {\em Proceedings of the VLDB Endowment}, vol.~13,
  no.~12, pp.~2382--2395, 2020.

\bibitem{papadomanolakis2007efficient}
S.~Papadomanolakis, D.~Dash, and A.~Ailamaki, ``Efficient use of the query
  optimizer for automated physical design,'' in {\em Proceedings of the 33rd
  International Conference on Very Large Databases (VLDB)}, pp.~1093--1104,
  2007.

\bibitem{dushnik1941partially}
B.~Dushnik and E.~W. Miller, ``Partially ordered sets,'' {\em American journal
  of mathematics}, vol.~63, no.~3, pp.~600--610, 1941.

\bibitem{appelgate1969ordinal}
H.~Appelgate, M.~Barr, J.~Beck, F.~Lawvere, F.~Linton, E.~Manes, M.~Tierney,
  F.~Ulmer, and F.~W. Lawvere, ``Ordinal sums and equational doctrines,'' in
  {\em Seminar on Triples and Categorical Homology Theory: ETH 1966/67},
  pp.~141--155, Springer, 1969.

\bibitem{costem}
MySQL Reference Manual 8.0, {\em Estimating Query Performance}, January 2023.
\newblock
  \url{https://dev.mysql.com/doc/refman/8.0/en/estimating-performance.html}.

\bibitem{martello1987algorithms}
S.~Martello and P.~Toth, ``Algorithms for knapsack problems,'' {\em
  North-Holland Mathematics Studies}, vol.~132, pp.~213--257, 1987.

\bibitem{chaudhuri2003factorizing}
S.~Chaudhuri, P.~Ganesan, and S.~Sarawagi, ``{Factorizing Complex Predicates in
  Queries to Exploit Indexes},'' in {\em Proceedings of the 2003 ACM SIGMOD
  International Conference on Management of Data}, pp.~361--372, 2003.

\bibitem{multirange}
MySQL Reference Manual 8.0, {\em Range Access Method for Multiple-Part
  Indexes}, October 2022.
\newblock
  \url{https://dev.mysql.com/doc/refman/8.0/en/range-optimization.html#range-access-multi-part}.

\bibitem{icp}
MySQL Reference Manual 8.0, {\em Index Condition Pushdown Optimization},
  October 2022.
\newblock
  \url{https://dev.mysql.com/doc/refman/8.0/en/index-condition-pushdown-optimization.htm}.

\bibitem{stjoin}
MySQL Reference Manual 8.0, {\em Join Clause}, October 2022.
\newblock \url{https://dev.mysql.com/doc/refman/8.0/en/join.html}.

\bibitem{toshihide1984optimal}
I.~Toshihide and K.~Tiko, ``On the optimal nesting order for computing
  n-relational joins,'' {\em ACM Transactions on Database Systems (TODS)},
  vol.~9, no.~3, pp.~482--502, 1984.

\bibitem{haritsa2005analyzing}
N.~Reddy and J.~R. Haritsa, ``{Analyzing Plan Diagrams of Database Query
  Optimizers},'' in {\em Proceedings of the 31st International Conference on
  Very Large Databases (VLDB)}, p.~1228–1239, 2005.

\bibitem{dash2011cophy}
D.~Dash, N.~Polyzotis, and A.~Ailamaki, ``Cophy: A scalable, portable, and
  interactive index advisor for large workloads,'' {\em arXiv preprint
  arXiv:1104.3214}, 2011.

\bibitem{dexter}
Dexter, {\em The automatic indexer for Postgres}, October 2022.
\newblock \url{https://github.com/ankane/dexter}.

\bibitem{whang1987index}
K.~Y. Whang, ``Index selection in relational databases,'' {\em Foundations of
  Data Organization}, pp.~487--500, 1987.

\bibitem{bruno2005automatic}
N.~Bruno and S.~Chaudhuri, ``{Automatic physical database tuning: A
  relaxation-based approach},'' in {\em Proceedings of the 2005 ACM SIGMOD
  International Conference on Management of Data}, pp.~227--238, 2005.

\bibitem{hypopg}
HypoPG, {\em Hypothetical Indexes for PostgreSQL}, October 2022.
\newblock \url{https://github.com/HypoPG/hypopg}.

\bibitem{innolim}
MySQL Reference Manual 8.0, {\em InnoDB Limits}, October 2022.
\newblock \url{https://dev.mysql.com/doc/refman/8.0/en/innodb-limits.html}.

\bibitem{zhou2022autoindex}
X.~Zhou, L.~Liu, W.~Li, L.~Jin, S.~Li, T.~Wang, and J.~Feng, ``{AutoIndex: An
  Incremental Index Management System for Dynamic Workloads},'' in {\em 2022
  IEEE 38th International Conference on Data Engineering (ICDE)},
  pp.~2196--2208, IEEE, 2022.

\bibitem{schnaitter2007line}
K.~Schnaitter, S.~Abiteboul, T.~Milo, and N.~Polyzotis, ``On-line index
  selection for shifting workloads,'' in {\em 2007 IEEE 23rd International
  Conference on Data Engineering Workshop (ICDEW)}, pp.~459--468, IEEE, 2007.

\bibitem{yadavflexiraft}
R.~Yadav and A.~Rahut, ``{FlexiRaft: Flexible Quorums with Raft},'' {\em The
  Conference on Innovative Data Systems Research (CIDR)}, 2023.

\bibitem{thein2014apache}
K.~M.~M. Thein, ``{Apache Kafka: Next generation distributed messaging
  system},'' {\em International Journal of Scientific Engineering and
  Technology Research}, vol.~3, no.~47, pp.~9478--9483, 2014.

\bibitem{wu2013predicting}
W.~Wu, Y.~Chi, S.~Zhu, J.~Tatemura, H.~Hacig{\"u}m{\"u}s, and J.~F. Naughton,
  ``Predicting query execution time: Are optimizer cost models really
  unusable?,'' in {\em 2013 IEEE 29th International Conference on Data
  Engineering (ICDE)}, pp.~1081--1092, IEEE, 2013.

\bibitem{skipscan}
MySQL Reference Manual 8.0, {\em Skip Scan Range Access Method}, October 2022.
\newblock
  \url{https://dev.mysql.com/doc/refman/8.0/en/range-optimization.html#range-access-skip-scan}.

\bibitem{imo}
MySQL Reference Manual 8.0, {\em Index Merge Optimization}, October 2022.
\newblock
  \url{https://dev.mysql.com/doc/refman/8.0/en/index-merge-optimization.html}.

\bibitem{skipscanbug}
MySQL Bug, {\em Skip Scan retrieves incorrect Result}, October 2022.
\newblock \url{https://bugs.mysql.com/bug.php?id=100253}.

\bibitem{imobug}
MySQL Bug, {\em Clustered primary key included in index merge may cause higher
  execution times}, October 2022.
\newblock \url{https://bugs.mysql.com/bug.php?id=80390}.

\bibitem{palermo1970quantitative}
F.~Palermo, ``A quantitative approach to the selection of secondary indexes,''
  {\em IBM Research, RJ}, vol.~730, 1970.

\bibitem{bruno2011configuration}
N.~Bruno and R.~V. Nehme, ``Configuration-parametric query optimization for
  physical design tuning,'' in {\em Proceedings of the 2008 ACM SIGMOD
  International Conference on Management of Data}, pp.~941--952, 2008.

\bibitem{wu2022budget}
W.~Wu, C.~Wang, T.~Siddiqui, J.~Wang, V.~Narasayya, S.~Chaudhuri, and P.~A.
  Bernstein, ``{Budget-aware Index Tuning with Reinforcement Learning},'' in
  {\em Proceedings of the 2022 ACM SIGMOD International Conference on
  Management of Data}, pp.~1528--1541, 2022.

\bibitem{borovica2012automated}
R.~Borovica, I.~Alagiannis, and A.~Ailamaki, ``Automated physical designers:
  what you see is (not) what you get,'' in {\em Proceedings of the Fifth
  International Workshop on Testing Database Systems}, pp.~1--6, 2012.

\bibitem{das2019automatically}
S.~Das, M.~Grbic, I.~Ilic, I.~Jovandic, A.~Jovanovic, V.~R. Narasayya,
  M.~Radulovic, M.~Stikic, G.~Xu, and S.~Chaudhuri, ``{Automatically indexing
  millions of databases in Microsoft Azure SQL database},'' in {\em Proceedings
  of the 2019 ACM SIGMOD International Conference on Management of Data},
  pp.~666--679, 2019.

\bibitem{caprara1995exact}
A.~Caprara, M.~Fischetti, and D.~Maio, ``Exact and approximate algorithms for
  the index selection problem in physical database design,'' {\em IEEE
  Transactions on Knowledge and Data Engineering (TKDE)}, vol.~7, no.~6,
  pp.~955--967, 1995.

\bibitem{caprara1996branch}
A.~Caprara and J.~Gonz{\'a}lez, ``A branch-and-cut algorithm for a
  generalization of the uncapacitated facility location problem,'' {\em Top},
  vol.~4, no.~1, pp.~135--163, 1996.

\bibitem{basu2016regularized}
D.~Basu, Q.~Lin, W.~Chen, H.~T. Vo, Z.~Yuan, P.~Senellart, and S.~Bressan,
  ``Regularized cost-model oblivious database tuning with reinforcement
  learning,'' in {\em Transactions on Large-Scale Data-and Knowledge-Centered
  Systems XXVIII}, pp.~96--132, Springer, 2016.

\bibitem{sharma2018case}
A.~Sharma, F.~M. Schuhknecht, and J.~Dittrich, ``The case for automatic
  database administration using deep reinforcement learning,'' {\em arXiv
  preprint arXiv:1801.05643}, 2018.

\bibitem{sadri2020online}
Z.~Sadri, L.~Gruenwald, and E.~Leal, ``Online index selection using deep
  reinforcement learning for a cluster database,'' in {\em 2020 IEEE 36th
  International Conference on Data Engineering Workshops (ICDEW)},
  pp.~158--161, IEEE, 2020.

\bibitem{perera2021dba}
R.~M. Perera, B.~Oetomo, B.~I. Rubinstein, and R.~Borovica-Gajic, ``{DBA
  bandits: Self-driving index tuning under ad-hoc, analytical workloads with
  safety guarantees},'' in {\em 2021 IEEE 37th International Conference on Data
  Engineering (ICDE)}, pp.~600--611, IEEE, 2021.

\bibitem{wang2021udo}
J.~Wang, I.~Trummer, and D.~Basu, ``{UDO: universal database optimization using
  reinforcement learning},'' {\em arXiv preprint arXiv:2104.01744}, 2021.

\bibitem{ding2019ai}
B.~Ding, S.~Das, R.~Marcus, W.~Wu, S.~Chaudhuri, and V.~R. Narasayya, ``{AI
  meets AI: Leveraging query executions to improve index recommendations},'' in
  {\em Proceedings of the 2019 ACM SIGMOD International Conference on
  Management of Data}, pp.~1241--1258, 2019.

\bibitem{hasan2020deep}
S.~Hasan, S.~Thirumuruganathan, J.~Augustine, N.~Koudas, and G.~Das, ``Deep
  learning models for selectivity estimation of multi-attribute queries,'' in
  {\em Proceedings of the 2020 ACM SIGMOD International Conference on
  Management of Data}, pp.~1035--1050, 2020.

\bibitem{guravannavar2012sort}
R.~Guravannavar, S.~Sudarshan, A.~A. Diwan, and C.~S. Babu, ``{Which Sort
  Orders Are Interesting?},'' {\em The VLDB Journal}, vol.~21, no.~1,
  pp.~145--–165, 2012.

\end{thebibliography}

\end{document}